\documentclass[wrr,draft]{agutex}

\usepackage{lineno}

\usepackage{graphicx}
\usepackage{amssymb}
\usepackage{amsmath}
\usepackage{psfrag}
\usepackage{tabularx}

\usepackage{ifthen}
\newboolean{@IsSubmitted}
\setboolean{@IsSubmitted}{false}
\setboolean{@IsSubmitted}{true}

\newboolean{@IsBWplots}
\setboolean{@IsBWplots}{true}
\setboolean{@IsBWplots}{false}

\newboolean{@IsAnalyticalDf}
\setboolean{@IsAnalyticalDf}{false}

\authorrunninghead{CHAPWANYA AND STOCKIE}

\titlerunninghead{SIMULATIONS OF GRAVITY-DRIVEN FINGERING}

\authoraddr{Michael Chapwanya, Department of Mathematics and Applied
  Mathematics, University of Pretoria, 0002 Pretoria, South Africa.
  (m.chapwanya@up.ac.za)}

\authoraddr{John M. Stockie, Department of Mathematics, Simon Fraser
  University, 8888 University Drive, Burnaby, British Columbia, V5A 1S6,
  Canada. (stockie@math.sfu.ca)}

\begin{document}

%
%


%
%


\title{Numerical Simulations of Gravity-Driven Fingering in Unsaturated
  Porous Media Using a Non-Equilibrium Model}

%
%


\author{Michael Chapwanya}
\affil{Department of Mathematics and Applied Mathematics, University of
  Pretoria, 0002 Pretoria, South Africa.}

\author{John M. Stockie}
\affil{Department of Mathematics, Simon Fraser University, 8888
  University Drive, Burnaby, British Columbia, V5A 1S6, Canada.} 

%
%

\begin{abstract}
  This is a computational study of gravity-driven fingering
  instabilities in unsaturated porous media.  The governing equations
  and corresponding numerical scheme are based on the work of Nieber et
  al. [Ch.~23 in \emph{Soil Water Repellency}, eds. C.~J.~Ritsema and
  L.~W.~Dekker, Elsevier, 2003] in which non-monotonic saturation
  profiles are obtained by supplementing the Richards equation with a
  non-equilibrium capillary pressure-saturation relationship, as well as
  including hysteretic effects.  The first part of the study takes an
  extensive look at the sensitivity of the finger solutions to certain
  key parameters in the model such as capillary shape parameter, initial
  saturation, and capillary relaxation coefficient.  The second part is
  a comparison to published experimental results that demonstrates the
  ability of the model to capture realistic fingering behaviour.
\end{abstract}

%
%


\begin{article}


\newcommand{\noi}{\noindent}
\newcommand{\scale}{^\ast}
\newcommand{\sat}{\theta}
\newcommand{\flux}{q}
\newcommand{\taus}{\tau\scale}
\newcommand{\psis}{\psi\scale}
\newcommand{\alphas}{\alpha\scale}
\newcommand{\Ps}{p\scale}
\newcommand{\Ks}{k\scale}
\newcommand{\xs}{x\scale}
\newcommand{\zs}{z\scale}
\newcommand{\nablas}{\nabla\scale}
\newcommand{\ts}{t\scale}
\newcommand{\sats}{\sat\scale}
\newcommand{\psiwe}{\psi_{\text{\itshape we}}}
\newcommand{\dtmax}{\Delta t_\text{\itshape max}}
\newcommand{\en}[1]{(\ref{eq:#1})}
\newcommand{\defeq}{\doteq}
\renewcommand{\defeq}{:=}
\newcommand{\porosity}{\phi}
\newcommand{\computer}{Mac Pro with 2$\times$3 GHz processor and 8GB RAM}
\newcommand{\MATLAB}{{\sc Matlab}}
\newcommand{\leavethisout}[1]{}
\newcommand{\units}[1]{$#1$}
\newcommand{\bunits}[1]{[$#1$]}
\newcommand{\myvec}[1]{\boldsymbol{#1}}
\newcommand{\myline}{\\}
\newcommand{\myfig}[3]{%
  \ifthenelse{\boolean{@IsBWplots}}{%
    \includegraphics[width=#1]{Figures/#3}}{%
    \includegraphics[width=#1]{Figures/#2}}%
}
\ifthenelse{\boolean{@IsSubmitted}}{
  \renewcommand{\myline}{\\[-0.4cm]}
  \renewcommand{\myfig}[3]{%
    \ifthenelse{\boolean{@IsBWplots}}{%
      \includegraphics[width=#1]{#3}}{%
      \includegraphics[width=#1]{#2}}}
}{}
\newcommand{\mylabelsize}{\footnotesize}
\newcommand{\mysmalllabelsize}{\scriptsize}
\newcommand{\replacewd}{%
  \psfrag{Depth, z}[cb][cB]{{\mylabelsize Depth, $z$}}%
  \psfrag{Width, x}[cB][cb]{{\mylabelsize Width, $x$}}}

\section{Introduction}
\label{sec:intro}

The transport of water and dissolved contaminants within the vadose zone
is extremely important in a wide range of natural and industrial
applications including protection of groundwater aquifers, irrigation, flood control,
and bioremediation, to name just a few.  Many of these applications exhibit
preferential flow in which gravitational, viscous or other forces
initiate instabilities that propagate as coherent finger-like
structures.  In fingered flow, water is able to bypass a significant
portion of the porous matrix and thus penetrate more rapidly than would
otherwise be possible for a uniform wetting front; as a result,
fingering can have a major impact on the transport of contaminants
carried by an infiltrating fluid.  A clear understanding of fingering
phenomena can therefore be essential in the study of certain
applications such as groundwater contamination.

We focus in this work on preferential flow that is driven by
gravitational forces arising from the difference in density between
invading water and displaced air.  The structure of a typical finger
consists of a nearly saturated ``tip'' at the leading edge, behind which
follows a ``tail'' region having a uniform and relatively low saturation
(see Fig.~\ref{fig:finger}).  As the finger penetrates into the soil,
the region immediately behind the tip drains somewhat causing pressure
to decrease and preventing the finger core from widening,\ 
thereby allowing the unstable finger to persist in time.  
Experimental studies have provided additional insight into the detailed
character of fingers and the physical mechanisms driving their
formation, beginning with the work of \citet{hill-parlange-1972} and
continuing to the present day with the work of authors such as
\citet{diment-watson-1985}, \citet{Glass-etal1990},
\citet{selker-etal-1992b}, \citet{lu-biggar-nielsen-1994},
\citet{Bauters-etal-2000b}, \citet{yao-hendrickx-2001},
\citet{WangJuryTuliKim2004} and \citet{DiCarlo2004}.

Various mathematical models have been developed to capture fingering
phenomena \citep{Philip1975,Parlange-Hill1976,DiCarlo2008} with many
based on applying the Richards equation (RE) in combination with
appropriate constitutive equations for soil properties.  Techniques of
linear stability analysis were applied to two-dimensional models by
\citet{Saffman-Taylor1958}, \citet{Chuoke-etal1959}, and
\citet{Parlange-Hill1976}, who derived stability criteria and
analytical predictions for quantities such as finger width and velocity.
\citet{Raats1973} postulated a criterion for stability which stated that
a wetting front is unstable if the velocity of the front increases with
depth; this is clearly satisfied for some layered media as well as for
water-repellent soils.\ \
Many analytical results have been compared to experiments by authors
such as \citet{Glass-Steenhuis-Parlange1989} and \citet{Wang-Feye1998},
who found that no single analytical formula is capable of capturing the
behaviour of the majority of soils.  Other modifications and
improvements to the theory have appeared more recently, such as
\citet{Wang-Feye1998} who modified the work of \citet{Parlange-Hill1976}
to include dependence on the water- and air-entry pressures. A
comprehensive review of stability results, including comparison to
experiments, can found in \citet{Rooij2000}.  There has also been a
great deal of recent effort on explaining gravity-driven fingering using
models based on conservation laws
\citep{Eliassi-Glass2002,Nieber-Sheshukov2003,Nieber-Dautov2005,cuetofelgueroso-juanes-2008,cuetofelgueroso-juanes-2009b}.
A recent paper by \citet{cuetofelgueroso-juanes-2009} presents the first
exhaustive stability analysis of a conservation law that leads to
fingering in unsaturated flow, and a follow-up study by the same authors
performs an extensive comparison to experiments as well as providing an
excellent review of the current literature
\citep{cuetofelgueroso-juanes-2009b}.

There has been a recent surge of interest in modelling fingering
instabilities using extensions of the RE model, such as the work of
\citet{Cuesta2000} and \citet{Cuesta-Hulshof2003} who analyze
non-monotonic travelling wave profiles that arise when dynamic capillary
effects are incorporated.  Both \citet{Eliassi-Glass2001a} and
\citet{Nieber-Dautov2005} identified a number of mechanisms that could
give rise to gravity-driven fingering, including non-monotonicity in
hydraulic properties, dynamic capillary effects and hysteresis.
\citet{Egorov-Dautov2003} provide an overview of the mathematical
formulation showing that Richards equation is unconditionally stable
even for heterogeneous media.  Furthermore, \citet{Nieber-Sheshukov2003}
claim that dynamic (or non-equilibrium) effects are sufficient to cause
formation of fingered flow, while persistence of fingers is dominated by
hysteresis.  In parallel with these developments, several novel
mathematical models have been developed which incorporate these and
other effects.  A number of authors have investigated the use of
non-equilibrium effects
\citep{Mitkov-etal-1998,Cuesta2000,Hassanizadeh2002,Helmig-etal-2007,Manthey-etal2008}
while others \citep{DiCarlo2008} have been inspired by non-monotonicity
to introduce extra terms in the RE that capture the
``hold-back-pile-up'' effect examined by \citet{Eliassi-Glass2003}.
Non-equilibrium effects have also been studied in the context of
two-phase flow by \citet{VanDuijn-Peletier-Pop2007}, who used an
extension of the Buckley-Leverett model to obtain non-monotonic profiles
with both infiltration and drainage fronts.
\citet{Sander-Glidewell-Norbury-2008} proposed a one-dimensional RE
model including hysteresis and non-equilibrium capillary terms, which is
very closely-related to the model studied in this paper.  Adding to the
controversy are experimental results such as \citet{dicarlo-2007} which
failed to find significant dynamic effects in gravity-driven
infiltration.  Notwithstanding the extensive literature on this subject,
many open questions remain about which governing equations and
constitutive relations are most appropriate for capturing preferential
flows.

We will focus on a specific model called the relaxation non-equilibrium
Richards equation (or RNERE)
\citep{Nieber-Sheshukov2003,Nieber-Dautov2005} which incorporates both
dynamic and hysteretic effects.  These authors developed an iterative
algorithm for integrating the governing equations numerically, and
showed that their method is capable of generating finger-like solutions.
The main drawback of this work was that it contained no concrete
comparisons to experimental results.  In this paper, we perform a more
extensive suite of numerical simulations with the RNERE model and
compare the results to published experimental studies.  We also carry
out a careful numerical convergence study and investigate the
sensitivity of the model to changes in physical parameters and
algorithmic aspects such as the choice of inter-block averaging for
hydraulic conductivity.  The results demonstrate that the RNERE model is
capable of reproducing realistic fingering flows for a wide range of
physically relevant parameters.

\section{Mathematical Model}
\label{sec:model}

We begin by presenting the governing equations for the RNERE model as
presented by \citet{Nieber-Sheshukov2003}, while at the same time
reviewing earlier work on fingered flow in porous media.  The RE is
written in mixed form as
\begin{gather}
\frac{\partial \sats}{\partial \ts}=\nablas \cdot \left( \Ks(\sats) 
  \nablas \psis\right) - \frac{\partial \Ks(\sats)}{\partial \zs},
  \label{eq:res}  
\end{gather}
where $\ts$ represents time \bunits{s}, $\sats$ is the volumetric water
content or saturation \bunits{m^3/m^3}, and $\psis$ is the water
pressure head \bunits{m}.  The asterisks are used here to indicate
dimensional quantities, and will be dropped shortly when the equations
are non-dimensionalized.  Hydraulic conductivity \bunits{m/s} is
denoted by $\Ks(\sats)$, which is assumed to be a given function of
water content in the case of unsaturated flow.  The spatial domain is
two-dimensional with coordinates $(\xs, \zs)$, where $\zs$ represents
the vertical direction and is measured positive downwards and $\xs$ is
measured horizontally.  This form of
the RE is called ``mixed'' because both saturation and pressure appear
as dependent variables, and it is preferred to both the $\sat$--based form
(which becomes singular when the flow is fully saturated) and the
$\psi$--based form (which leads to large mass conservation errors when
discretized) \citep{Celia1990}.

\begin{figure}[bthp]
  \centering
  \myfig{0.4\textwidth}{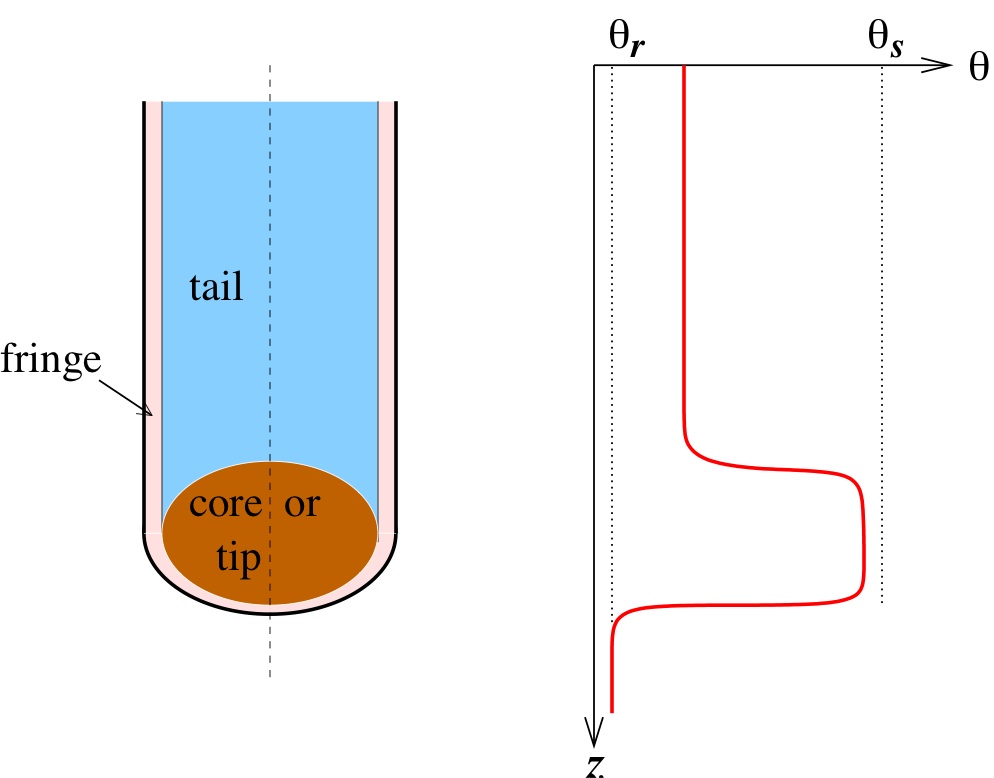}{fingercore.jpg}
  \caption{Left: A typical finger propagating into a porous medium
    having an initially uniform saturation $\sat_r$.  At the leading
    edge of the finger is a well-defined ``core'' or ``tip'' region with
    water content close to the saturated value $\sat_s$.  Behind the tip
    lies a ``tail'' region having a nearly constant intermediate value
    of water content.  Right: The corresponding saturation profile along
    the central axis of the finger.}
  \label{fig:finger}
\end{figure}

Many models of flow in porous media combine the RE with an equilibrium
constitutive relation of the form $\psis=\Ps(\sats)$, which assumes that
transport properties relax instantaneously to their equilibrium values
as water content varies during a wetting or drying process.  This is a
reasonable approximation under certain circumstances; however, there is
now evidence from both laboratory experiments \citep{DiCarlo2004} and
stability analyses \citep{Nieber-Dautov2005} that suggests
the RE by itself is unable to capture the non-monotonic profiles
observed in fingering instabilities and so it lacks some critical 
physical mechanism.\ \
An illustration of a typical solution profile is shown in
Fig.~\ref{fig:finger}, wherein a downward-propagating finger is led by a
nearly-saturated ``tip'' region that leaves behind it a ``tail'' region
having a lower water content.\ \
An earlier attempt at simulating fingered flow using the equilibrium RE
was made by \citet{Nieber1996} who incorporated hysteretic effects and
found that fingers only appeared when a downwind-weighted mean was used
for hydraulic conductivity.  \citet{Eliassi-Glass2001a} concluded
that the finger-like profiles obtained in these simulations did not
represent solutions of the actual model equations, but rather were 
numerical artifacts arising from local truncation errors due to the
particular choice of downwind mean.

Based on physical arguments, \citet{Hassanizadeh1993} advocated that a
non-equilibrium version of the capillary pressure relationship should be
employed in situations where the relaxation time is comparable to other
time scales in the flow.  This work inspired
\citet{Nieber-Sheshukov2003} to propose their RNERE model in which the
RE was supplemented by a relaxation equation of the form
\begin{gather}
  \psis - \Ps(\sats) = 
  \frac{\taus}{\rho g}\frac{\partial \sats}{\partial \ts}, 
  \label{eq:newrelaxs} 
\end{gather}  
where $\Ps(\sats)$ represents the equilibrium water pressure head
\bunits{m}, $\rho$ is the density of water \bunits{kg/m^3}, $g$ is the
gravitational acceleration \bunits{m/s^2}, and
$\taus=\taus(\psis,\sats)>0$ is a suitably-chosen capillary relaxation
function \bunits{kg/m\, s}.  They presented numerical simulations of
finger-like instabilities and concluded that a non-equilibrium effect is
sufficient to initiative the instabilities and that hysteresis is
necessary to sustain the fingers once formed.  In most cases, $\taus$ is
assumed either to be a constant
\citep{Hassanizadeh2002,Manthey-etal2008} or else a separable function
of dynamic capillary pressure and water content
\citep{Duijn2004,DiCarlo2005,Nieber-Dautov2005}, although the proper
choice of functional form for the relaxation function remains an open
question.  We note that Eq.~\en{newrelaxs} should be viewed as an equation
for the dynamic capillary pressure $\psis$ rather than an evolution
equation for $\sats$; indeed, when \en{newrelaxs} is substituted into
Eq.~\en{res}, the resulting PDE takes the form of a third-order
evolution equation for $\sats$ which is of pseudo-parabolic
type~\citep{King-Cuesta2006}.

Before proceeding further, we briefly mention several other attempts at
incorporating non-equilibrium effects into the RE in more general
contexts not directly related to fingering.  \citet{Mitkov-etal-1998}
took a phase-field model for solidification and adapted it to porous
media flow; their model contains a phenomenological term in which the
constants have no direct relationship to the physics.
\citet{Barenblatt-etal-2003} suggested an alternate approach in which
dynamic effects are incorporated into both capillary pressure and
hydraulic conductivity through an ``effective saturation'' variable.
Three further variants of the RE called the hypo-diffusive, hyperbolic
and mixed forms were proposed with an aim to reproducing the
``hold-back-pile-up'' effect observed in experiments
\citep{Eliassi-Glass2003,DiCarlo2008}.  Analytical and numerical results
suggest that many of these approaches show promise, but the proper
choice of model remains an open question.

To complete the mathematical description of the RNERE model equations
\en{res} and \en{newrelaxs}, the equilibrium pressure $\Ps$
and hydraulic conductivity $\Ks$ must be specified as functions of water
content.  These quantities are customarily expressed in terms of the
normalized water content
\begin{gather}
  \sat = \frac{\sats - \sat_r}{\sat_s - \sat_r},
  \label{eq:sat} 
\end{gather}
where $\sat_s$ and $\sat_r$ are the saturated and residual (or
irreducible) water contents, respectively; $\sat$ is commonly referred
to as the effective saturation or simply saturation.  In a partially
saturated porous medium, the saturation variable satisfies $0\leqslant
\sat_r \leqslant \sats \leqslant \sat_s \leqslant \porosity$, where
$\porosity$ represents the porosity, so that $\sat$ always lies between
0 and 1.  We adopt the widely-used van~Genuchten--Mualem relationships
for $\Ps(\sat)$ and $\Ks(\sat)$ \citep{Genuchten1980}, which are
monotonic functions containing several empirical fitting parameters that
are used to fit with experimental data for a variety of soil and rock
types.  Saturation and pressure are related at equilibrium by
\begin{gather}
  \sat = \left( 1 + \alphas_\ell|\Ps|^{n_\ell} \right)^{-m_\ell},
  \label{eq:sat_pres}
\end{gather}
where $n_\ell$ and $m_\ell=1-1/n_\ell$ are parameters that govern the
shape of the capillary curves.  

In practice, $\sat$ is a hysteretic function wherein the inverse
capillary length $\alphas_\ell$ \bunits{m^{-1}} and shape parameter
$n_\ell$ differ depending on whether the current state is evolving along
the main wetting curve ($\ell=w$) or main drying curve ($\ell=d$).  The
corresponding hydraulic conductivity is given by
\begin{gather}
  \Ks(\sat) = k_o \sqrt{\sat} \left[ 1-(1-\sat^{1/{m_\ell}})^{m_\ell}
  \right]^2, 
  \label{eq:ksats}
\end{gather}
where $k_o$ represents the fully saturated value (at $\sat=1$).  The RE
\en{res} becomes degenerate when $\Ks(0)=0$, although we never have to
deal explicitly with this issue because we always choose a value of
initial saturation $\sat_i$ greater than the residual value $\sat_r$.
Furthermore, we have not encountered values of the reduced saturation in
excess of 1 (for which the conductivity is undefined in Eq.~\en{ksats})
and so the computed solution remains within the physical limits
$0\leqslant\sat\leqslant 1$.  Details of the precise form of the
hysteresis model to be used will be provided later in
Section~\ref{sec:algorithm}.

\subsection{Non-Dimensionalization and Choice of $\tau$}

In this section, the equations are reduced to dimensionless form using
the transformations
\begin{gather}
  x = \alphas_w \xs, \qquad z = \alphas_w \zs, \qquad
  \alpha_\ell = \alphas_\ell/\alphas_w,
  \nonumber \\
  \psi= \alphas_w\psis, \qquad p = \alphas_w \Ps,  \qquad
  k = \Ks/k_o, 
  \label{eq:nondim} \\
  t = \alphas_w k_o\ts/(\sat_s-\sat_r), 
  \nonumber
\end{gather}
where $(\alphas_w)^{-1}$ (the reciprocal of the van Genuchten parameter
for the main wetting curve) has been used as the natural length scale.
The governing equations \en{res}, \en{newrelaxs}, \en{sat_pres} and
\en{ksats} then reduce to
\begin{align}
  \frac{\partial \sat}{\partial t} &= \nabla \cdot \left( k(\sat) \,
    \nabla \psi \right)- \frac{\partial k(\sat)}{\partial z},
  \label{eq:re} \\ 
  \psi &= p + \overline{\tau}(\psi,\sat)\,\frac{\partial \sat}{\partial t}, 
  \label{eq:relax}\\
  \sat &= S_\ell(p) \defeq \left( 1 + \alpha_\ell |p|^{n_\ell}
  \right)^{-m_\ell}, 
  \label{eq:sat_pre}\\
  k(\sat) &= K_\ell(\sat) \defeq \sqrt{\sat} \left(
    1-(1-\sat^{1/m_\ell})^{m_\ell}\right)^2, 
  \label{eq:ksat}
\end{align}
where the dimensionless capillary relaxation function is 
\begin{gather*}
  \overline{\tau}(\psi,\sat) = \frac{(\alphas_w)^2 k_o}{\rho g} \,
  \taus(\psis,\sats),
\end{gather*}
and $S_\ell(p)$ and $K_\ell(\sat)$ represent the hysteretic constitutive
relations (which depend on the wetting/drying state $\ell=w, d$).  It
will prove convenient when describing the numerical algorithm
to recast the time-derivative in Eq.~\en{relax} in terms of the
equilibrium capillary pressure as
\begin{gather}
   \tau(\psi,\sat) \, \frac{\partial p}{\partial t} = \psi - p, 
\label{eq:newrelax}
\end{gather} 
where $\tau(\psi,\sat) = \overline{\tau}(\psi,\sat) \, (d\sat/dp)$.

Following \citet{Nieber-Dautov2005} (who was motivated by the
experiments of \citet{Selker-Parlange1992}) we assume that $\tau$ is a
function of $\psi$ only
\begin{gather}
  \tau(\psi)= \tau_o [\psi - \psi_o]^{\gamma}_+,
  \label{eq:tau-psi}
\end{gather}
where $\tau_o$, $\psi_o$ and $\gamma$ are constants and $[\cdot]_+ =
\max(\cdot,0)$.  Nonetheless, the appropriate choice of functional
dependence for $\tau$ on the state variables remains an open question.
Various other functional forms have been proposed by \citet{DiCarlo2005}
and \citet{Sander-Glidewell-Norbury-2008}, all of which we find leads to
similar fingering patterns provided that $\tau\rightarrow 0$ as $\psi
\rightarrow \psi_o$ and that the magnitude of the relaxation parameter
is comparable.  On the other hand, if $\tau$ is taken to be a constant
then no fingers were observed in our simulations, hence suggesting that
it is essential to have a solution-dependent $\tau$.

\subsection{Boundary and Initial Conditions}
\label{sec:bcs}

\begin{figure}[bthp]
  \centering  
  \psfrag{q1=0}[cc][bc]{{\mylabelsize $\flux_1 = 0$}}
  \psfrag{q2=qi}[cb][Bc]{{\mylabelsize $\flux_2=\flux_i$}}
  \psfrag{q2=qi + q~}[Bc][Bc]{{\mylabelsize $\flux_2=\flux_i+\widetilde{\flux}(x)$}}
  \psfrag{infiltration zone}[cb][Bc]{{\mylabelsize infiltration zone ($d_i$)}}
  \psfrag{x}[cb][Bc]{{\mylabelsize $x$}}
  \psfrag{z}[cb][Bc]{{\mylabelsize $z$}}
  \myfig{0.3\textwidth}{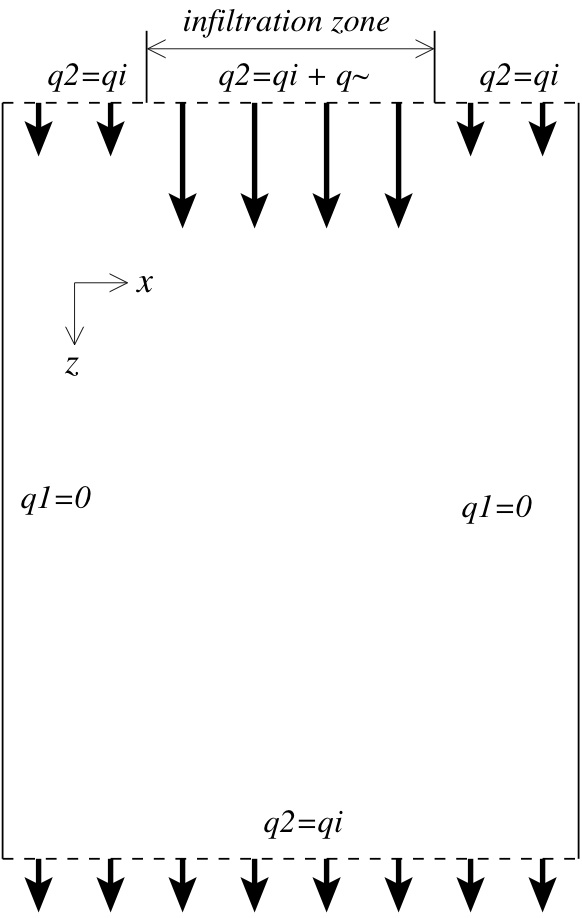}{flowdomain.jpg}
  \caption{The computational domain with width $L$ and height $H$. Zero
    flux conditions are imposed on side boundaries and a background flux
    of $\flux_i$ on the top and bottom boundaries.  Finger formation is
    driven by an additional infiltration flux $\widetilde{\flux}$
    applied along a portion of the top boundary having width $d_i$.}
  \label{fig:domain} 
\end{figure}

Inspired by the geometry most commonly employed in experimental studies,
we consider a two-dimensional rectangular domain as shown in
Fig.~\ref{fig:domain} that has width $L$ and height $H$ (both dimensions
having been non-dimensionalized by scaling with $\alphas_w$ like the
other lengths in Eq.~\en{nondim}).\ \
The initial saturation is assumed constant throughout the domain,
$\sat(x,z,0) = \sat_i$.  No-flux conditions are imposed along side
boundaries and specified inflow (outflow) conditions are given along the
top (bottom) boundaries, both of which we express in terms of the
dimensionless flux variable
\begin{gather}
  \myvec{\flux} = (\flux_1,\flux_2) = -k(\sat) \, \nabla (\psi-z),
  \label{eq:v-darcy}
\end{gather}
where $\flux_1$ and $\flux_2$ represent the components of the flux
vector. 
This last equation is a statement of Darcy's law for unsaturated flow
and is rescaled according to $\myvec{\flux}=\myvec{\flux}\scale/k_o$,
where the original physical flux $\myvec{\flux}\scale$ has units of
\units{m/s}.  The fluxes on the left, right, and bottom boundaries are
given respectively by
\begin{gather} 
  \flux_1(0,z,t) = 0, \quad
  \flux_1(L,z,t) = 0, \quad
  \flux_2(x,H,t) = \flux_i,\label{eq:bc1}
\end{gather}
where $\flux_i$ represents a constant gravity-driven background flux
that corresponds to the initial saturation $\sat_i$.  The specification
of such a background flux is essential when the porous medium is not
air-dried, such as is the case for most naturally occurring soils.  In
order to drive the formation of fingers, the flux along the top boundary
is specified by some constant background flux $\flux_i$ plus an
infiltration flux $\widetilde{\flux}$ that is imposed along a
strip of width $0\leqslant d_i\leqslant L$,
\begin{gather} 
  \flux_2(x,0,t)=\left\{
    \begin{array}{ll}
      \flux_i + \widetilde{\flux}(x), & \text{if $|2x-L|\leqslant d_i$,} \\
      \flux_i, & \text{otherwise.}
    \end{array}
  \right. \label{eq:bc2}
\end{gather}
Following \citet{Nieber-Sheshukov2003}, this infiltration flux is
written as the sum of an average value $\flux_s$ plus a small sinusoidal
perturbation
\begin{gather}
  \widetilde{\flux}(x) = \flux_s + \flux_s\eta\cos\left( 
    \frac{\pi f}{d_i}( 2x-L+d_i ) \right), 
  \label{eq:bc3}
\end{gather}
where $\eta$ represents the amplitude of the perturbation and $2\pi
f/d_i$ the frequency (for $f$ a positive integer).  We note that the
size and number of fingers actually observed in simulations is
relatively insensitive to the choice of perturbation parameters $\eta$
and $f$.
 
We close with a brief mention of a common result on stability of
gravity-driven vertical infiltration flow, wherein viscous forces tend
to stabilize the flow while gravitational forces are the destabilizing
influence.  Extensive work on stability has been reported by many
authors, including \citet{Philip1975}, \citet{Parlange-Hill1976},
\citet{Wang-Feye1998} and \citet{Rooij2000}.  It is known that unstable
flow will occur if the hydraulic conductivity increases with depth,
which translates into a requirement that the inflow at the top boundary
satisfies $0<\flux_i+\flux_s<1$.

\section{Solution Algorithm}
\label{sec:algorithm}

We next describe the algorithm developed by \citet{Nieber-Sheshukov2003}
for solving the RNERE problem in Eqs.~\en{re} and
\en{sat_pre}--\en{newrelax}, which is an iterative strategy that employs
a finite volume spatial discretization in space and a semi-implicit
time-stepping scheme.  The domain is divided into an $N_x\times N_z$
rectangular grid, with cell dimensions $\Delta x=L/N_x$ and $\Delta
z=H/N_z$ in the $x$-- and $z$--directions respectively.  The discrete
saturation $\sat_{i,j}$ approximates the solution at cell centers
$\left( (i-1/2)\Delta x, (j-1/2)\Delta z \right)$, and similarly for the
pressure head $\psi_{i,j}$.  Employing an implicit backward Euler
discretization for the time derivative in Eqs.~\en{re} and \en{newrelax}
and centered second-order differences in space, the discrete equations
become
\begin{multline}
  \frac{ \sat_{i,j} - \widehat{\sat}_{i,j} }{\Delta t} = \\
  \frac{1}{\Delta x}\left( k_{i+1/2,j}
    \frac{\psi_{i+1,j}-\psi_{i,j}}{\Delta x} - k_{i-1/2,j}
    \frac{\psi_{i,j}-\psi_{i-1,j}}{\Delta x}\right) + \\
  \frac{1}{\Delta z}\left( k_{i,j+1/2}
    \frac{\psi_{i,j+1}-\psi_{i,j}}{\Delta z} - k_{i,j-1/2}
    \frac{\psi_{i,j}-\psi_{i,j-1}}{\Delta z}\right) - \\
  \frac{k_{i,j+1/2}-k_{i,j-1/2}}{\Delta z}, 
  \label{eq:discRE} 
\end{multline}
and
\begin{gather} 
  \tau(\psi_{i,j}) \, \frac{{p}_{i,j}-\widehat{p}_{i,j}}{\Delta t} =
  \psi_{i,j}-p_{i,j},
  \label{eq:discw}  
\end{gather}
for $i=1, 2, \dots, N_x$ and $j=1, 2, \dots, N_z$.  The time step is
denoted by $\Delta t$ and the ``hat'' notation $\widehat{\sat}$ and
$\widehat{p}$ refers to a solution value at the previous time step.

It is worth mentioning that although an upwind difference might normally
be advocated for the convective (gravitational) term in Eq.~\en{discRE},
we have chosen to use a centered difference for consistency.  In
practical computations, we observe no difference between an upwind or
centered difference treatment of the $\partial k/\partial z$ term
because capillary effects dominate in this problem.

In contrast with the cell-centered saturation and pressure head
unknowns, the hydraulic conductivity values $k_{i\pm 1/2,j}$ and
$k_{i,j\pm 1/2}$ are located at cell edges.  Since the conductivity
depends on saturation which is not available at cell edges, it must be
approximated using some weighted mean of nearby values of saturation;
the specific choice of averaging method will be considered in detail in
Section~\ref{sec:sens-k}.  The capillary relaxation function in
Eq.~\en{tau-psi} is replaced by the regularized function
$\tau_\delta(\psi)=\tau_o\,\max\left((\psi-\psi_o)^\gamma, \delta\right)$,
where the cut-off parameter $0<\delta\ll 1$ prevents $\tau$ from
becoming zero and hence avoids a singularity in Eq.~\en{discw}.

The difference stencils in Eq.~\en{discRE} involve values of pressure
head $\psi_{i,j}$ for $i=0,N_x+1$ and $j=0,N_y+1$, located at points
lying one-half grid cell outside the physical domain.  These
``fictitious values'' are eliminated using the flux boundary conditions
\en{bc1}--\en{bc3} as follows.  First, the flux is discretized along
cell edges, with the $\flux_1$-component (on side boundaries $i=0,
    N_x$) being approximated by
\begin{gather}
  \flux_{1;i+1/2,j} = -k_{i+1/2,j}\left(
    \frac{\psi_{i+1,j}-\psi_{i,j}}{\Delta x} \right),
  \label{eq:discBC1}\\
  \intertext{and the $\flux_2$-component (along horizontal boundaries
    $j=0,N_z$) by}  
  \flux_{2;i,j+1/2} = -k_{i,j+1/2}
  \left(\frac{\psi_{i,j+1}-\psi_{i,j}}{\Delta z} - 1\right). 
  \label{eq:discBC2}
\end{gather}
Then the boundary conditions for $\flux_1$ and $\flux_2$ are used to
express fictitious point values in terms of known values of pressure
head at interior points, which can then be used in Eq.~\en{discRE}.

We now describe the iterative scheme for solving the nonlinear system
\en{discRE}, which can be written more succinctly in matrix-vector form
as
\begin{gather}  
  \mathbb{A}\, \Pi+\frac{\Theta - \widehat{\Theta} }{\Delta t} = 0,
  \label{eq:it1} 
\end{gather}
where $\Theta$ and $\Pi$ are vectors containing the discrete
approximations of $\sat_{i,j}$ and $(\psi-z)_{i,j}$ respectively, and
$\mathbb{A}=\mathbb{A}(\Theta)$ is a symmetric pentadiagonal matrix
whose entries are nonlinear functions of saturation.
\citet{Nieber-Sheshukov2003} did not base their iterative solution
strategy directly on Eq.~\en{it1} because the matrix $\mathbb{A}$ is not
positive definite; instead, they proposed the following modified
iteration
\begin{gather} 
  (\mathbb{A}^{\nu+1}+\mathbb{D}^{\nu+1})\Pi^{\nu+1} =
  \mathbb{D}^{\nu+1}\Pi^{\nu} - 
  \frac{\Theta^{\nu+1}- \widehat{\Theta} }{\Delta t},
  \label{eq:iteration} 
\end{gather} 
where $\nu$ represents the iteration number and $\mathbb{D}$ is a
diagonal matrix whose entries are given by
\begin{gather}
  \mathbb{D} = \frac{1}{\Delta t}\frac{\partial\Theta}{\partial\Psi}
  = \frac{S'(P)}{\Delta t} \, \frac{d}{d\Psi}\left( \frac{\Psi\Delta t
      +\tau(\Psi)\widehat{P}}{\tau(\Psi) + \Delta t}
  \right) .
  \label{eq:dmat}
\end{gather}
This approach has the advantage that the iteration matrix
$(\mathbb{A+D})$ is both symmetric and positive definite and thus has
much better convergence properties.  We now outline the iterative
procedure within each time step, assuming that each iteration begins with
$\nu=0$, $\Theta^0=\widehat{\Theta}$, $\Psi^0=\widehat{\Psi}$ and
$P^0=\widehat{P}$:
\begin{description}
\item[Step 1.] Solve the relaxation equation 
  \begin{gather*}
    \tau(\Psi^\nu)\, \frac{P^{\nu+1}-\widehat{P}}{\Delta t} = \Psi^\nu -
    P^{\nu+1}
  \end{gather*}
  for $P^{\nu+1}$.
\item[Step 2.] Update the saturation using $\Theta^{\nu+1}=S(P^{\nu+1})$.    
\item[Step 3.] Evaluate the matrices $\mathbb{A}$ and $\mathbb{D}$ at
  $\Theta=\Theta^{\nu+1}$, $P=P^{\nu+1}$ and $\Psi=\Psi^{\nu}$.  Then
  solve the linear system \en{iteration} for $\Pi^{\nu+1}$ and let
  $\Psi^{\nu+1}=\Pi^{\nu+1}+z$.
\item[Step 4.] If the current solution satisfies the convergence
  criterion $\|\Pi^{\nu+1}-\Pi^\nu\|_2/\|\Pi^\nu\|_2 < 10^{-6}$, then
  stop.  Otherwise, increment $\nu$ and return to Step 1.
\end{description}

Following \citet{Eliassi-Glass2001a}, we employ a variable time step
which is initialized to $\Delta t=10^{-4}$ and then replaced at the end
of each time step by $\min(1.05 \Delta t,\, \dtmax)$, where the maximum
allowable step is given by the CFL-like condition $\dtmax = 0.1 \,
\min(\Delta x, \Delta z) / \flux_s$.  This approach minimizes initial
start-up errors by taking a relatively small time step initially, which
then increases gradually to $\dtmax$.  The algorithm just described is
implemented in \MATLAB\ and uses the built-in preconditioned conjugate
gradient solver {\tt pcg} to invert the linear system in Step 3.

We stress that this is only one possible choice of algorithm and that
many other strategies have been proposed for solving the coupled system
of equations for saturation and capillary pressure.  For example,
\citet{Cuesta2003} and \citet{Cuesta-Pop2009} have analyzed a number of
algorithms (including the one described above) in the context of the
Burgers equation, supplemented by dynamic capillary effects.

An important aspect of our RNERE model is the singularities that occur when
$\sat=0$ (where the iteration matrix $(\mathbb{A+D})$ fails to be
positive definite) and $\sat=1$ (where the derivative of the hydraulic
conductivity $k^\prime(\sat)$ becomes unbounded).  A number of methods
have been proposed in the literature (e.g.,
\citet{starke-2000,pop-2002}) to regularize coefficients in the
governing equations in order to avoid these singularities.  We have not
made use of any such regularization here because in practice, we find
that the computed saturation never reaches the limiting values of 0 or
1.  Nonetheless, it may be worthwhile in future to consider implementing
such a regularization approach to improve the efficiency and robustness
of the algorithm in cases where conditions approach the saturated and
unsaturated limits.

\subsection{Implementation of Hysteresis}
\label{sec:hysteresis}

An integral component of the RNERE algorithm is the specification of the
hysteretic state, which is potentially different at every point in the
domain and depends on the local saturation and wetting history.  We have
chosen to implement a closed-loop hysteresis model described by
\citet{Scott1983} and implemented by \citet{Eliassi-Glass2003} wherein
all curves have the same values of residual and saturated water content
(0 and 1 respectively in our dimensionless variables).  We have also
taken the shape parameters for the wetting and drying curves to be
constant and equal ($n\defeq n_w=n_d$), so that the various curves differ only
in their value of $\alpha_\ell$ (although in general, the values of
$n_\ell$ should also depend on the current hysteretic state).

Following \citet{Eliassi-Glass2003}, the main drying curve is written
$\sat = S_d(p)$ while the scanning drying curves are given by the scaled
equation
\begin{gather}  
  \sat = \frac{\overline{\sat} S_d (p)}{S_d(\overline{p})}, \label{eq:sdc}
\end{gather}
where $\overline{\sat}$ and $\overline{p}$ denote respectively the
saturation and pressure reversal points along the previous wetting
curve.  Similarly, the main wetting curve is $\sat=S_w(p)$ while the
scanning wetting curves are written
\begin{gather}  
  \sat = \overline{\sat}_{\text{\emph{rev}}} + 
  ({1-\overline{\sat}_{\text{\emph{rev}}}}) S_w(p),\label{eq:swc1}
\end{gather}
where
\begin{gather}  
  \overline{\sat}_{\text{\emph{rev}}} = 
  \frac{\overline{\sat} - S_w(\overline{p})}{1 - S_w(\overline{p})},  
  \label{eq:swc2} 
\end{gather}
and $\overline{\sat}$ and $\overline{p}$ are the reversal points along the
previous drying curve.   

The main drying and wetting curves are unique, while the scanning curves
differ depending on the reversal points which are determined as follows.
At each point in space, we maintain the current state (wet or dry) as
well as the previous reversal point $(\overline{\sat}, \overline{p})$.
To avoid problems with convergence in the iterative scheme, the
hysteretic state is updated only at the end of each time step and not
within a $\nu$-iteration.  To detect a reversal point along a local
drying or wetting curve, we test whether the time rate of change of
saturation has reversed sign between the current ($k$) and previous
($k-1$) time steps, which is equivalent to checking that $\Delta
\sat_{i,j}^{k}\cdot \Delta \sat_{i,j}^{k-1} < 0$ where $\Delta
\sat_{i,j}^{k} = \sat_{i,j}^{k}-\sat_{i,j}^{k-1}$.  To avoid spurious
wet/dry oscillations between successive time steps, we impose the
additional constraint that $\left| \Delta \sat_{i,j}^k \right| >
\varepsilon$, where $\varepsilon$ is a reversal threshold.  If both of
these criteria are met, then a flow reversal has occurred and the
current state is switched to either wetting (if $\Delta
\sat_{i,j}^k\geqslant 0$) or drying (if $\Delta \sat_{i,j}^k < 0$), and
the current values for the reversal point
$(\overline{\sat},\overline{p})$ are updated.  The appropriate scanning
curve -- either Eq.~\en{sdc} or \en{swc1}--\en{swc2} -- is then used to
determine the capillary pressure as a function of saturation.

\section{Numerical Simulations}
\label{sec:num}

To investigate the relevance of the proposed model and the accuracy and
efficiency of the numerical algorithm, we consider a ``base case''
corresponding to a 14/20 grade sand studied experimentally by
\citet{Glass-Steenhuis-Parlange1989}.  The parameters listed in
Table~\ref{tab:params} are taken directly from their paper, with the
exception of $n$, $\sat_i$ and $\alpha_w$, whose values are justified in
Sections~\ref{sec:sens-n}--\ref{sec:sens-tau}.

The computational domain is taken to be a rectangle of width $L=14$ and
height $H=35$ in dimensionless units, where $L$ is chosen slightly
larger than the actual infiltration width $d_i=10.5$ used in experiments
in order to minimize boundary effects.  Unless otherwise noted, the
domain is discretized using a uniform grid having $N_x=201$ and $N_z=401$
points in the $x$ and $z$ directions respectively.  All simulations were
performed on a \computer, with a typical run requiring approximately $2$
hours of computation time.
\begin{table}[bthp]
  \caption{``Base case'' parameters given in SI units.  With the
    exception of $n$, $\sat_i$ and $\alpha_w$, these values are taken
    from \citet{Glass-Steenhuis-Parlange1989}.} 
  \label{tab:params} 
  \begin{center}
    \begin{tabular}{clcc}
      \tableline
      Symbol & Description & Value & Units \\
      \tableline
      $n$        & Capillary shape parameter & 12     & --              \\
      $\alpha_w$ & Inverse capillary length  & 35     & \units{m^{-1}}  \\
      $\sat_i$   & Initial water content     & 0.01   & \units{m^3/m^3} \\
      $\sat_s$   & Saturated water content   & 0.42   & \units{m^3/m^3} \\
      $\sat_r$   & Residual water content    & 0.075  & \units{m^3/m^3} \\
      $k_o$      & Saturated conductivity    & 0.063  & \units{m/s}     \\
      $K$        & Permeability              & $6.5\times10^{-10}$ & \units{m^2}\\   
      $\psiwe$   & Water entry pressure      &$-0.023$& \units{m}       \\
      \tableline
    \end{tabular}
  \end{center}
\end{table}
\begin{table}[bthp]
  \caption{Dimensionless parameter values: (a) ``base case'' and Glass
    et al. comparisons; (b)--(c) modifications to base values for 
    simulations indicated.}     
  \label{tab:params-nond} 
  \begin{center}
    \begin{tabular}{clc}
      \tableline
      Symbol & Description & Value \\
      \tableline
      \multicolumn{3}{p{0.95\columnwidth}}{{\itshape (a) Parameters from
          \citet{Glass-Steenhuis-Parlange1989} -- ``base case'':}}\myline
      $n$        & Capillary shape parameter & 12.0 \myline
      $\alpha_w$ & Inverse capillary length (wetting)& 1.0 \myline
      $\alpha_d$ & Inverse capillary length (drying) & 0.5 \myline
      $\sat_i$   & Initial water content     & 0.01\myline
      $\tau_o$   & Relaxation coefficient    & 0.1 \myline
      $\gamma$   & Relaxation exponent       & 1   \myline
      $\psi_o$   & Relaxation parameter      & 0   \myline
      $\varepsilon$& Hysteretic reversal criterion & $10^{-10}$\myline
      $\delta$   & Relaxation cut-off        & 0.04\myline
      $\flux_i$  & Background flux           & $3.3\times 10^{-6}$\myline
      $\flux_s$  & Infiltration flux         & 0.14\myline
      $\eta$     & Perturbation amplitude    & 0.01\myline
      $f$        & Perturbation frequency    & 5   \myline
      $d_i$      & Infiltration source width & 10.5\myline
      $t_{\text{\itshape end}}$ & End time   & 77  \myline  
      $H$        & Domain height             & 35  \myline
      $L$        & Domain width              & 14  \\[0.2cm]
      \multicolumn{3}{p{0.95\columnwidth}}{{\itshape (b) Modifications for
          \citet{Nieber-Sheshukov2003} (Figs.~\ref{fig:nieber2}
          \&\ \ref{fig:nieber1}):}}\myline
      $n$        & Capillary shape parameter & 7.0 \myline
      $\sat_i$   & Initial water content     & 0.1 \myline
      $\tau_o$   & Relaxation coefficient    & 5.0 \myline
      $\flux_s$  & Infiltration flux         & 0.2 \myline
      $t_{\text{\itshape end}}$ & End time   & 96  \myline  
      $H$        & Domain height             & 60  \myline
      $L$        & Domain width              & 30  \\[0.2cm]
      \multicolumn{3}{p{0.95\columnwidth}}{{\itshape (c) Modifications for
          \citet[Tab.~1]{DiCarlo2004} for 20/30 and 30/40 sands
          (Fig.~\ref{fig:2030-3040}):}}\myline   
      $n$        & Capillary shape parameter & 6.23 / 10.0 \myline
      $\sat_i$   & Initial water content     & 0.001       \myline
      $H$        & Domain height             & 7.08 / 6.92 \myline
      $L$        & Domain width              & 0\tablenotemark{a} \\
      \tableline
    \end{tabular}
    \tablenotetext{a}{\citeauthor{DiCarlo2004}'s experimental soil
      columns measure less than one finger width in diameter, so that
      the flow is essentially one-dimensional; we perform ``quasi-1D''
      simulations by taking only two grid points in the $x$--direction.}
  \end{center}
\end{table}

\begin{figure}
  \centering
  \replacewd
  \myfig{0.4\textwidth}{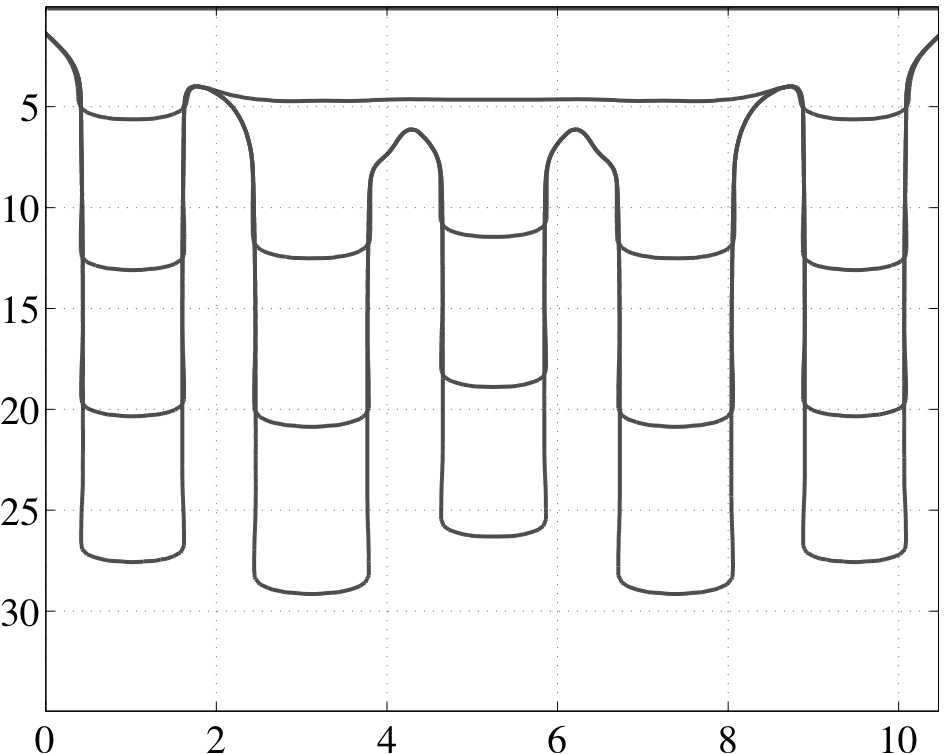}{base_only.jpg}
  \caption{Saturation contours for the base case, plotted at 
    four equally-spaced time intervals between $t=0$ and
    $t_{\text{\itshape end}}=77$.} 
  \label{fig:base-contour}
\end{figure}

\begin{figure}
  \centering
  \replacewd
  \myfig{0.5\textwidth}{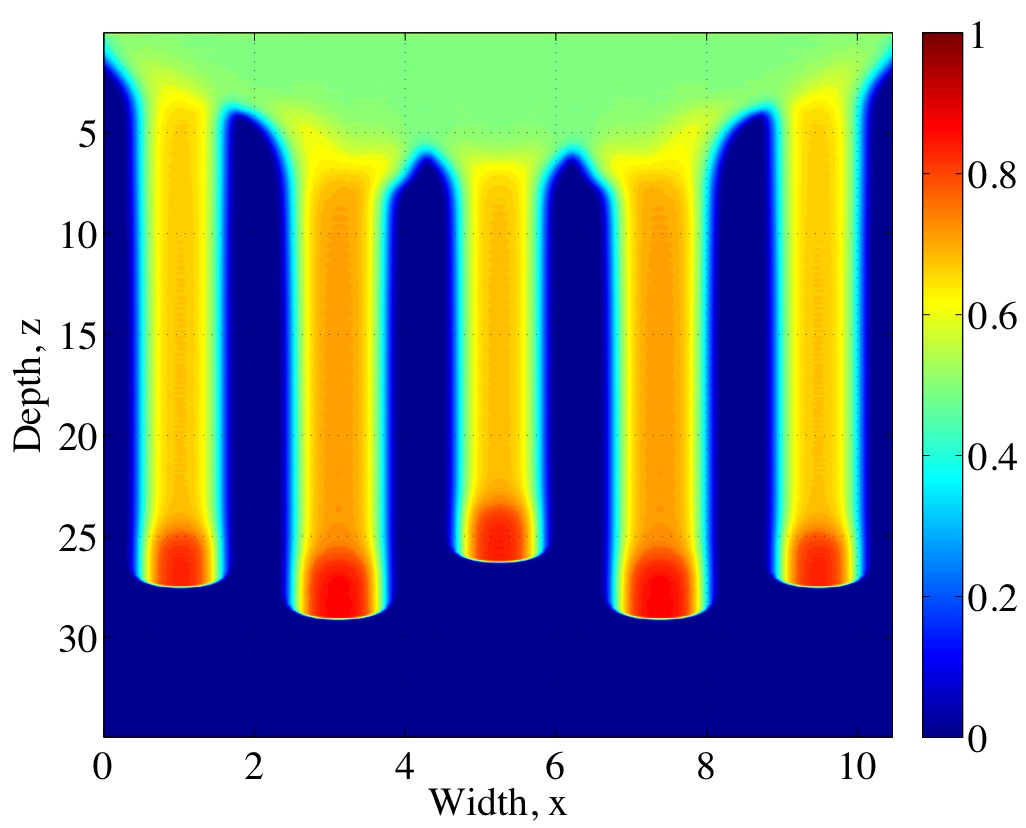}{base_only_bw.jpg}
  \caption{Saturation map for the base case at time
    $t_{\text{\itshape end}}=77$.}
  \label{fig:base-map}
\end{figure}

A sample computation with the base case parameters is shown in
Fig.~\ref{fig:base-contour}, which depicts the progression of the
wetting front at a sequence of equally-spaced times.  In these plots (as
well as the other plots and tables that follow) all quantities are
expressed in dimensionless form.  The plotted contours of saturation
correspond to a value of $\sat$ equal to 25\%\ of the finger tail
saturation, which we have found gives a good representation of the
finger size and shape.  The structure of the individual fingers is seen
more clearly in the saturation map given in Fig.~\ref{fig:base-map},
where each concentrated finger tip is followed by a tail region
of roughly constant saturation.  The ``capillary fringe'' region,
depicted schematically in Fig.~\ref{fig:finger}, is evident as a narrow
zone of rapid saturation change surrounding each finger.\ \
The finger tip and tail are evident in this plot and the shape
of each finger is in qualitative agreement with the generic profile
sketched in Fig.~\ref{fig:finger}.

\begin{figure}
  \centering
  \replacewd
  \begin{tabular}{m{0.2cm}m{2cm}m{2cm}m{2cm}}
    & (a) & (b) & (c)
  \end{tabular}\\
  \myfig{0.4\textwidth}{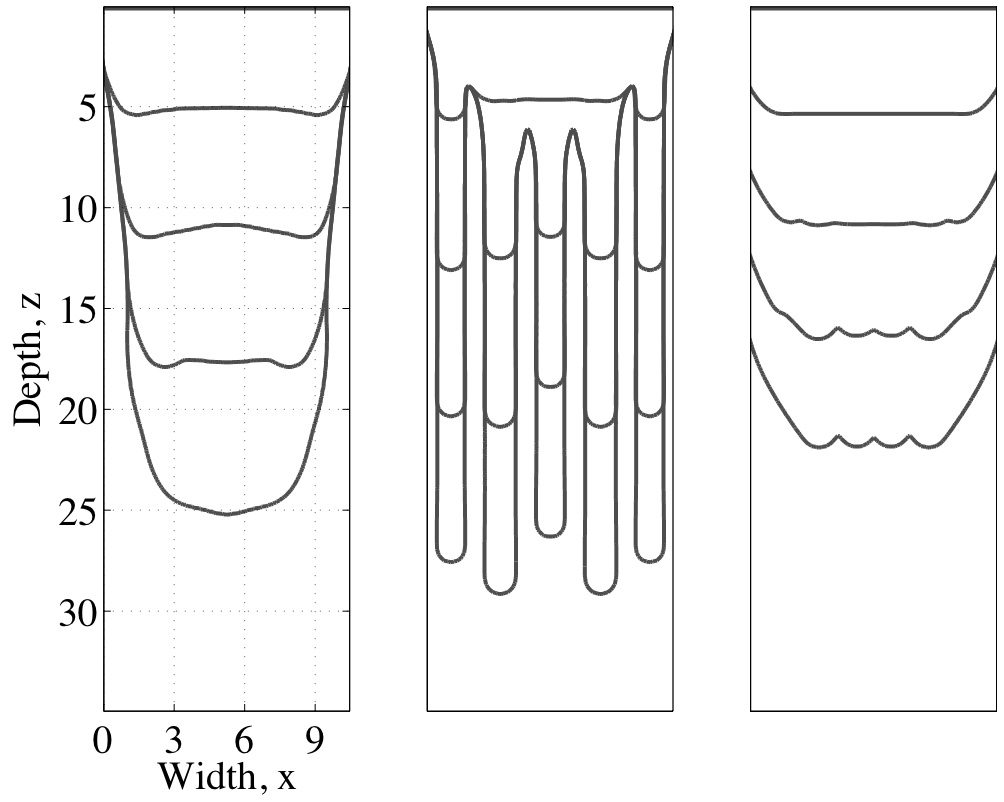}{method_sumplots.jpg}
  \caption{The effect of hysteresis and dynamic effects on the wetting
    front, for the base case: (a) with hysteresis only; (b) base case,
    with both hysteresis and dynamic effects; (c) with dynamic effects
    only.}
  \label{fig:method-compare}
\end{figure}

To illustrate the importance of dynamic and hysteretic effects in the
RNERE model, we show in Fig.~\ref{fig:method-compare} how the wetting
front differs when either of these effects is left out.  When the
dynamic term is omitted from the saturation equation (see
Fig.~\ref{fig:method-compare}(a)) fingering instabilities clearly fail
to be initiated.  Conversely, when hysteretic effects are left out (see
Fig.~\ref{fig:method-compare}(c)) protrusions begin to form at the
wetting front but they never actually develop into full-blown fingers.
These observations are consistent with the claim of
\citet{Nieber-Sheshukov2003} that dynamic capillary effects are
responsible for the initiation of fingering instabilities, while
hysteresis is required to sustain the fingers in time.

In the following sections, we present an extensive suite of numerical
simulations that address the following four issues:
\begin{itemize}
\item choosing an appropriate mean for the estimation of inter-block
  hydraulic conductivity values;
\item determining the dependence of the numerical solution on grid
  resolution, and comparing to previously published simulations;
\item measuring the sensitivity of the solution to certain key
  parameters: shape parameter ($n_\ell$), dynamic relaxation
  coefficient ($\tau$) and initial saturation ($\sat_i$); and
\item comparing simulated results to previously published experimental
  data.
\end{itemize}
\noindent
In addition to providing plots of saturation plots, we will also report
quantities such as finger width ($d_f$), finger velocity ($v_f$), number
of fingers ($N_f$), and average volume flow rate through each finger
($Q_f$), all of which vary depending on the value of the infiltration
flux $\flux_s$.  When multiple fingers are present and a specific
quantity varies from finger to finger, we report the average value over
all fully-developed fingers.

\subsection{Choice of Mean for Inter-Block Conductivity}
\label{sec:sens-k}

As mentioned in Section~\ref{sec:algorithm}, the discrete equations
require values of hydraulic conductivity at cell edges ($k_{i\pm 1/2,j}$
and $k_{i,j\pm 1/2}$) whereas the values of saturation on which $k$
depends are defined at cell centers; therefore, some form of averaging
is usually necessary.  It is well known that discrete approximations of
the RE can be very sensitive to the choice of inter-block averaging used
for hydraulic conductivity \citep{Belfort2005}.  A number of different
approaches have been advocated in the literature, for instance using
arithmetic \citep{Vandam-Feddes-2000}, geometric \citep{Haverkamp1979},
harmonic \citep{Das1994}, and Darcian-weighted means
\citep{Warrick1991}.  \citet{cardwell-parsons-1945} showed that the
effective permeability for a heterogeneous porous medium must lie
somewhere between the harmonic and arithmetic mean values.  Furthermore,
\citet{warren-price-1961} used Monte Carlo simulations of random media
to show that the expected value of conductivity for a heterogeneous
system is given by the geometric mean.

Of particular interest in this paper is the case where the conductivity
undergoes large variation between grid cells owing to the presence of
sharp wetting fronts at finger boundaries.  Although a straightforward
analytical argument indicates that the harmonic mean is the appropriate
mean to use in such situations in 1D \citep{gutjahr-etal-1978}, this is
not the case in higher dimensions where many computational studies
indicate that the harmonic average yields results that are inferior to
those using other averaging methods
\citep{Haverkamp1979,Belfort2005,pinales-etal-2005}.  Extensive
comparisons have been drawn using measures such as resolution and
stability of the wetting fronts, sensitivity to grid refinement, and
robustness over a wide range of soil types and physical parameters.
There remains a significant degree of controversy over which averaging
procedure is best in practice, and to date no single mean has been found
to be superior in all circumstances.

\leavethisout{ 
  The schemes that smear the front lead to an artificial stabilization
  of the flow while those that sharpen the front can lead to artificial
  destabilization of the flow, \citet{Zaidel1992}.
  \citet{Haverkamp1979} compared nine different averaging methods for
  initially dry soils and concluded that the GM generates little
  weighting error. In fact, the geometric mean is less affected by
  changing the space increments. \citet{Das1994} compared eight
  different averaging methods and their results were inconclusive though
  the arithmetic mean and the harmonic mean gave better results.
  \citet{Zaidel1992} observed that the arithmetic mean (for initially
  dry soil) is most accurate for the van Genuchten parameter $n\geqslant 2$
  though it significantly smears the front. This smearing of the front
  was also observed in the work of \citetg{Warrick1991} with the GM
  giving sharp fronts. Recently, \citet{Belfort2005} found that the GM
  can significantly improve results even for large grid spacing.  
}

In this section, we compare results using the arithmetic and geometric
means for hydraulic conductivity, which we have found are the most
common means utilized in computations.  Values of conductivity along
vertical cell edges are determined as follows
\begin{align*}
  \text{Arithmetic mean:} & \quad 
  k_{i\pm 1/2,j} = \frac{1}{2}(k_{i\pm 1,j}+k_{i,j}),\\
  \text{Geometric mean:}  &\quad 
  k_{i\pm 1/2,j} = \sqrt{k_{i\pm 1,j}k_{i,j}}\, , 
\end{align*}
with similar formulas for $k_{i,j\pm 1/2}$ along horizontal cell edges.\
\ 
Problem parameters are taken from simulations presented by
\citet{Nieber-Sheshukov2003}, which are identical to those for our base
case described earlier, except for a few differences indicated in
Table~\ref{tab:params-nond}(b).  The same set of parameters will also be
considered in the following two sections.  Our model is identical to
\citeauthor{Nieber-Sheshukov2003}'s, except for a slight difference in
the implementation of capillary hysteresis.

\begin{figure}[hbtp]
  \centering  
  \psfrag{X}{}
  \psfrag{201}{}
  \psfrag{401}{}
  \psfrag{801}{}
  \psfrag{101cells}{}
  \psfrag{201cells}{}
  \psfrag{401cells}{}
  \replacewd
  \myfig{0.4\textwidth}{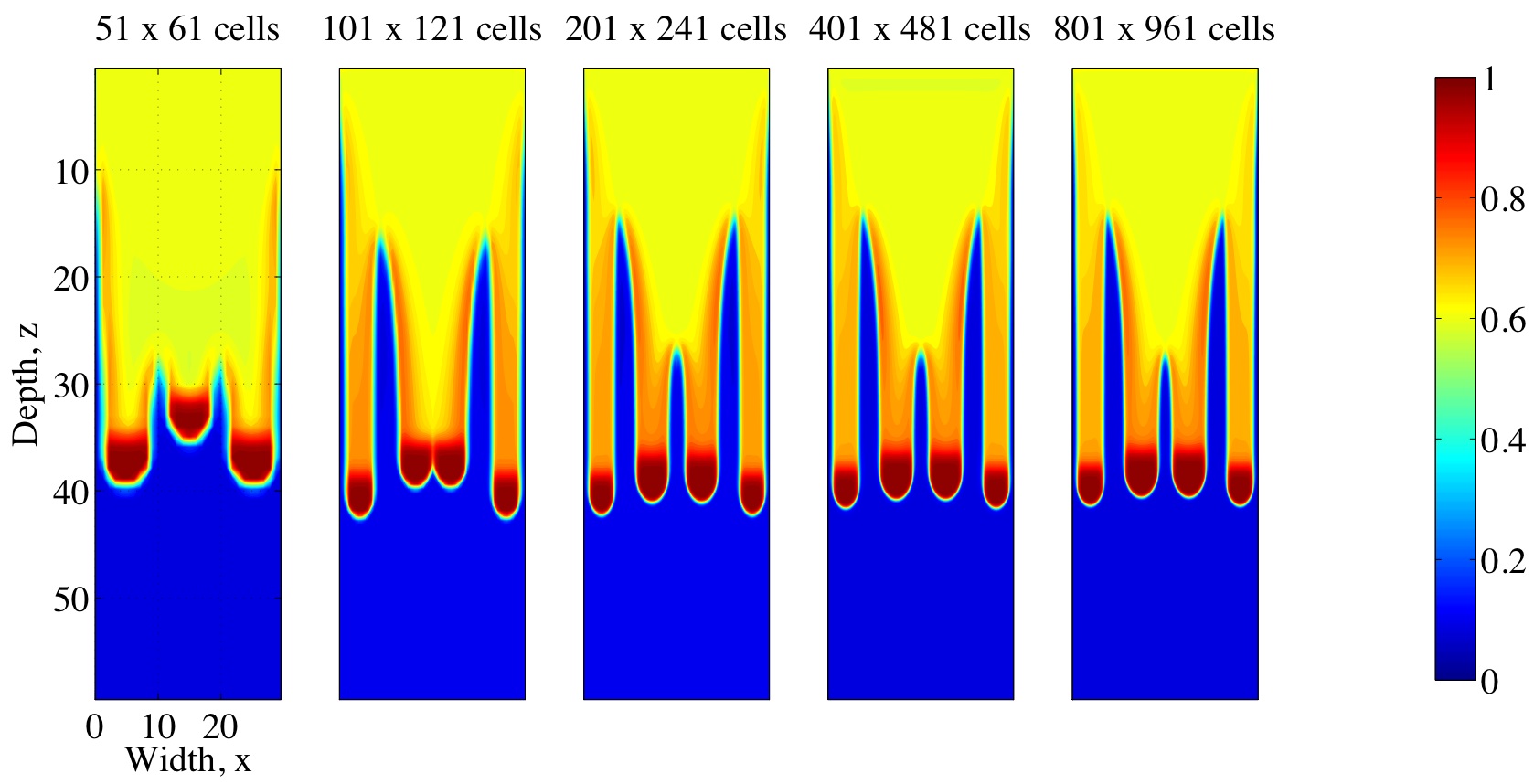}{nieber_gm_sumplots_bw.jpg}
  \caption{Saturation maps corresponding to the geometric mean for
    five different grid resolutions, using parameters from
    \citet{Nieber-Sheshukov2003}.} 
  \label{fig:gm}
\end{figure}

\begin{figure}[hbtp]
  \centering  
  \psfrag{X}{}
  \psfrag{201}{}
  \psfrag{401}{}
  \psfrag{801}{}
  \psfrag{101cells}{}
  \psfrag{201cells}{}
  \psfrag{401cells}{}
  \replacewd
  \myfig{0.4\textwidth}{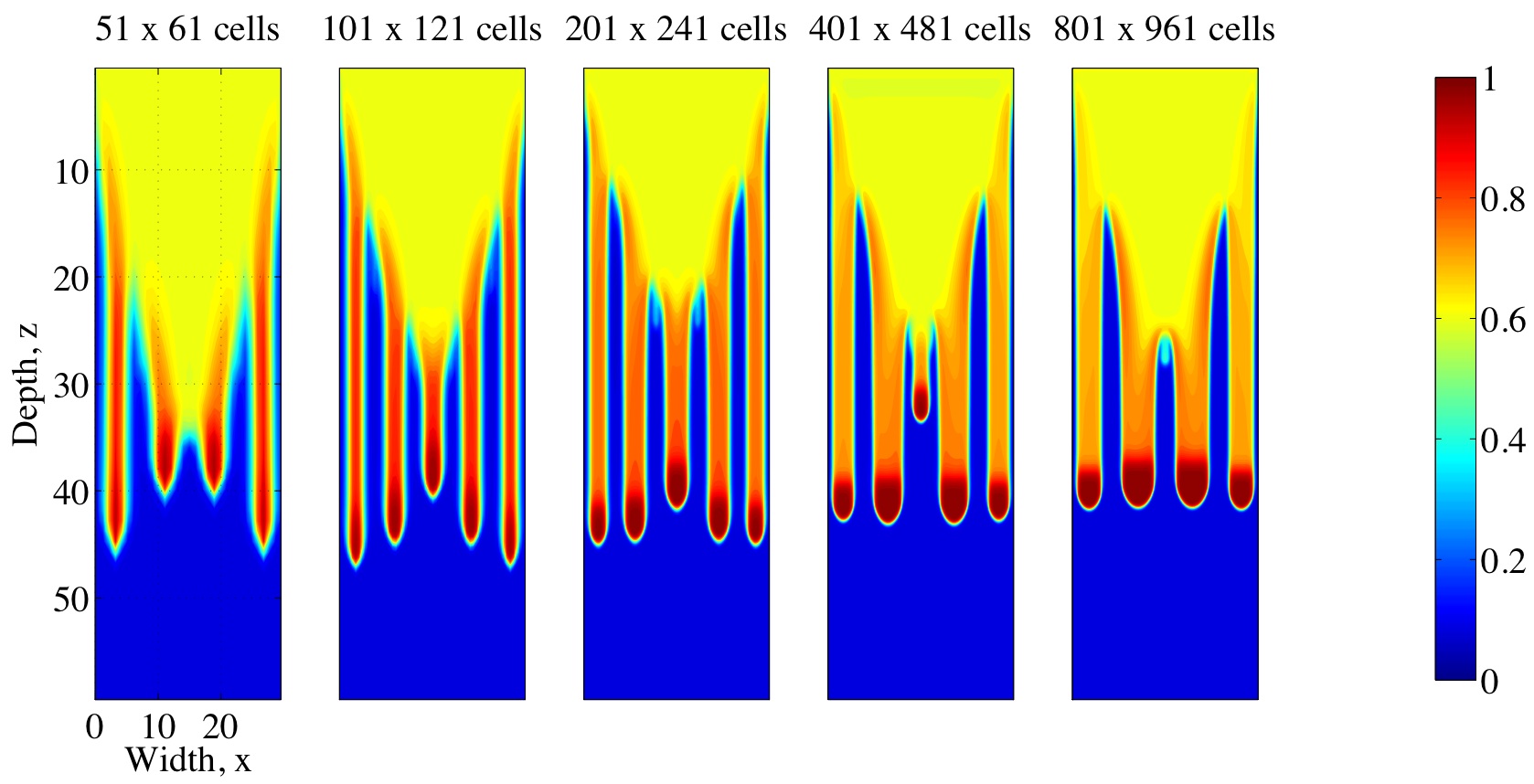}{nieber_am_sumplots_bw.jpg}
  \caption{Saturation maps corresponding to the arithmetic mean for
    five different grid resolutions, using parameters from
    \citet{Nieber-Sheshukov2003}.}
  \label{fig:am}
\end{figure}

The results for the geometric mean are shown in Fig.~\ref{fig:gm} for
five different choices of grid resolution ($51\times 61$, $101\times
121$, $201\times 241$, $401\times 481$ and $801\times 961$), and the
solution has clearly converged to a wetting profile with 4 well-defined
fingers even on a $201\times 241$ grid.  In contrast, Fig.~\ref{fig:am}
demonstrates that simulations with the arithmetic mean converge more
slowly, and the results on the finest grid are still not fully
converged.  These simulations are consistent with those of
\citet{Zaidel1992}, who found that use of the arithmetic mean introduces
excessive smearing in wetting fronts and underestimates saturation
values relative to the geometric mean.  Based on these results, we
conclude that the geometric mean is superior for the problem under
consideration, which is also consistent with the a number of previous
studies \citep{Haverkamp1979,hornung-messing-1983,Belfort2005}.
Consequently, we have chosen to apply the geometric mean in all
remaining computations in this paper.


\leavethisout{ In addition to the two averaging methods presented in
  this paper, we also considered the harmonic mean (although the results
  not shown here).  By observing the saturation along the central axis
  of a chosen finger, we observed that the harmonic mean overestimates
  the maximum finger tip saturation while the arithmetic mean
  underestimates the finger tip saturation for lower grid resolutions.
}

\subsection{Grid Refinement Study}
\label{sec:grid}

To ensure that the numerical solution does converge with the expected
second order accuracy, simulations were performed on a sequence of
successively refined grids of size $101 \times 121$ to $801\times 961$
cells.  The latter represents the finest resolution possible owing to
memory restrictions on the computing equipment readily available to us.
The grid resolution and physical parameters in this case were chosen to
correspond to the numerical simulations of \citet{Nieber-Sheshukov2003}.

\begin{figure}[hbtp]
  \centering
  \replacewd
  \psfrag{121 by 101 cells}[c][c]{{\scriptsize $101 \times 121$ cells}} 
  \psfrag{241 by 201 cells}[c][c]{{\scriptsize $201 \times 241$ cells}} 
  \psfrag{481 by 401 cells}[c][c]{{\scriptsize $401 \times 481$ cells}} 
  \psfrag{961 by 801 cells}[c][c]{{\scriptsize $801 \times 961$ cells}} 
  \myfig{0.4\textwidth}{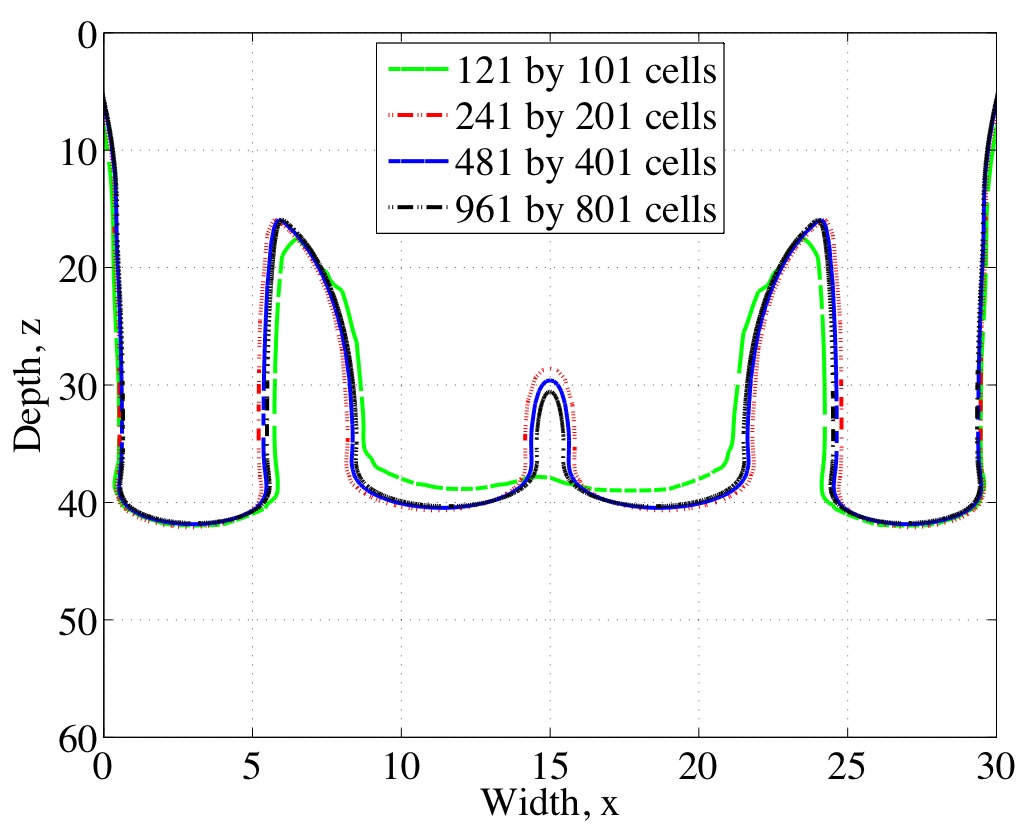}{nieberg_contour_plot.jpg}
  \caption{Saturation contours corresponding to the different grid
    resolutions used in the convergence study.} 
  \label{fig:nieber2}
\end{figure}

The solution on the finest grid is treated as the ``exact solution'' and
the error is estimated using the $\ell_2$ norm of the difference between
exact and computed values of saturation.  The resulting absolute errors
are summarized in Table~\ref{tab:refinement} from which it is clear that
the solution converges as the grid is refined; furthermore, the order of
convergence is close to the expected value of 2.  Fig.~\ref{fig:nieber2}
depicts saturation contours corresponding to various grid refinement
levels and clearly demonstrates the convergence of the numerical
solution.
\begin{table}[hbtp]
  \caption{Grid refinement study, where the order of accuracy is
    estimated as the base-2 logarithm of the ratio of successive
    errors.  The ``exact'' solution corresponds to an $801\times 961$
    grid computation.}  
  \label{tab:refinement}
  \begin{tabular}{cccc}
    \hline 
    No. of cells ($N_x\times N_z$) & $\ell_2-$ error & Ratio & Order \\
    \hline
    $ 60 \times 51$  & 10.59 & 3.52 & 1.82 \\
    $121 \times 101$ & 3.01  & 3.82 & 1.93 \\
    $241 \times 201$ & 0.79  & 4.31 & 2.11 \\
    $481 \times 401$ & 0.18  &  --  & --   \\
    \hline
  \end{tabular}
\end{table}

\subsection{Comparison with Nieber et al.'s Computations}
\label{sec:compare-nieber}

We now investigate the effects of grid resolution in more detail by way
of a direct comparison with simulation results reported by
\citet{Nieber-Sheshukov2003}.  We focus on their second set of
simulations (c.f., their Fig.~8), using different values of the
infiltration width ($d_i=1$, 5, 10.5, 15, 20, 25 and 30) and grid
resolutions of $101 \times 121$ and $201 \times 241$.  Our results are
summarized in Fig.~\ref{fig:nieber1} from which we observe a close match
with \citet[Fig.~8]{Nieber-Sheshukov2003} in terms of both number of
fingers and saturation levels.  There are slight differences between
finger widths and velocities, but we attribute these to our alternative
implementation of capillary hysteresis.  We emphasize the discrepancies
between our high and low resolution results for values of $d_i=10.5$, 15
and 30, which stresses the importance of using a sufficiently resolved
grid in these fingering computations.  It is particularly important to
use a higher grid resolution in situations such as $d_i=30$ that are
close to a ``transitional phase'' where in this case the solution
exhibits somewhere between three and four fingers.  We also stress the
importance of performing a careful convergence analysis as part of any
numerical study of fingering to ensure that the features being simulated
are a true representation of actual fingering instabilities of the
governing equations.

\leavethisout{
  In Fig.~\ref{fig:nieber1}, a relaxation coefficient $\tau_o=5.0$ was
  used throughout the simulations. This value is found to be very high
  compared to the best fit observed in the base case simulations. Thus
  the pile-up-effect (in the finger tip) is very high and mass
  conservation requires that less number of fingers be formed. 
}
\ifthenelse{\boolean{@IsSubmitted}}{\begin{figure}}%
  {\begin{figure*}}
  \psfrag{i}{}
  \psfrag{ =1}[ct][lb]{{\mylabelsize  (a) $d_i=1$\quad}}
  \psfrag{ =5}[ct][lb]{{\mylabelsize  (b) $d_i=5$\qquad}}
  \psfrag{ =10}[ct][lb]{{\mylabelsize (c) $d_i=10.5$\quad}}
  \psfrag{ =15}[ct][lb]{{\mylabelsize (d) $d_i=15$\quad}}
  \psfrag{ =20}[ct][lb]{{\mylabelsize (e) $d_i=20$\quad}}
  \psfrag{ =25}[ct][lb]{{\mylabelsize (f) $d_i=25$\quad}}
  \psfrag{ =30}[ct][lb]{{\mylabelsize (g) $d_i=30$\quad}}
  \psfrag{\(a\) d}{}
  \psfrag{\(b\) d}{}
  \psfrag{\(c\) d}{}
  \psfrag{\(d\) d}{}
  \psfrag{\(e\) d}{}
  \psfrag{\(f\) d}{}
  \psfrag{\(g\) d}{}
  \replacewd
  \def\ph{\phantom{age }}
  \def\pph{\phantom{th}}
  \newbox\dothis
  \setbox\dothis=\vbox to0pt{\vskip-1pt\hsize=11.5pc\centering
    Error\vss}
  \centering
  \myfig{0.85\textwidth}{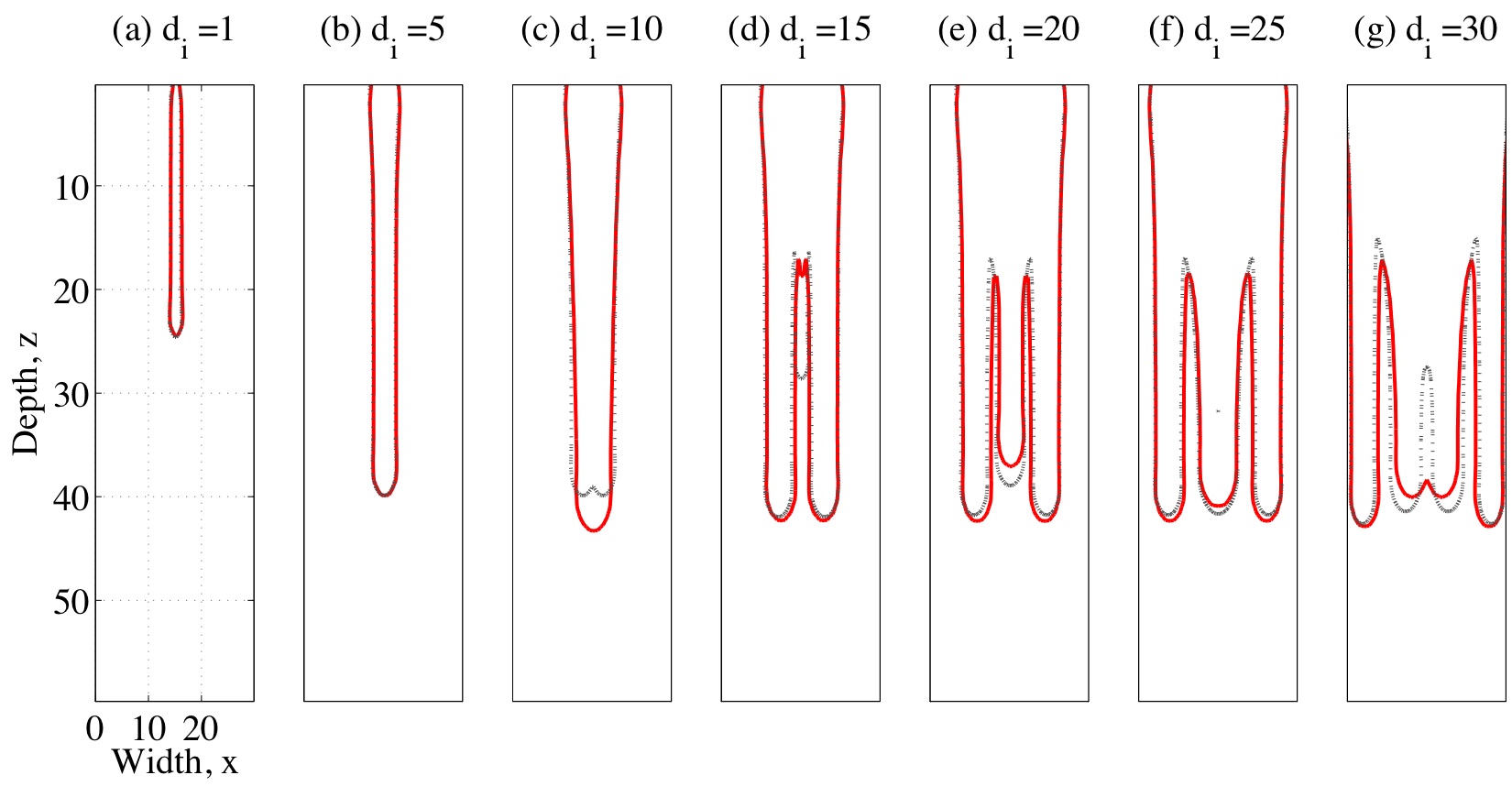}{nieberall.jpg}
  \caption{Comparison of fingers for various values of the infiltration
    width $d_i$, based on parameters from
    \citet{Nieber-Sheshukov2003}.  The solid (red) saturation contours
    correspond to solutions on a $101\times 121$ grid, while the broken 
    (black) contours are for a $201 \times 241$ grid.}
  \label{fig:nieber1}
\ifthenelse{\boolean{@IsSubmitted}}{\end{figure}}%
  {\end{figure*}}

\subsection{Sensitivity to Capillary Shape Parameter, $n$}
\label{sec:sens-n}

In the next three sections, we switch to the base case and investigate
the sensitivity of the solution to changes in a number of important
parameters.  No value is provided by
\citet{Glass-Steenhuis-Parlange1989} for the capillary shape parameter
$n$ appearing in the van Genuchten--Mualem relationships \en{sat_pre}
and \en{ksat}, and so we look for guidance in related experimental
studies on sandy soils.  The values for $n$ reported in the literature
exhibit significant variability, lying anywhere between 3 and 20 even
for porous media having similar coarseness and wettability (e.g.,
\citet{schroth-etal-1996,nieber-etal-2000}).\ \
The larger values of $n$ typically correspond to water-repellent soils
in which fingers are more likely to form, while smaller values indicate
a reduced tendency to generate fingering instabilities.  Consequently,
it is important in any modelling study of fingering to understand the
effect of changes in $n$ on the character of the solution.

\begin{figure}[hbtp]
  \centering  
  \psfrag{n = 4}{}
  \psfrag{n = 8}{}
  \psfrag{n = 12}{}
  \psfrag{n = 15}{}
  \replacewd
  {\mylabelsize
    \begin{tabular}{cccc}
      \;\; (a) $n=4$ & (b) $n=8$ &
      (c) $n=12^\dag$ & (d) $n=15$
    \end{tabular}\\
  }
  \myfig{0.4\textwidth}{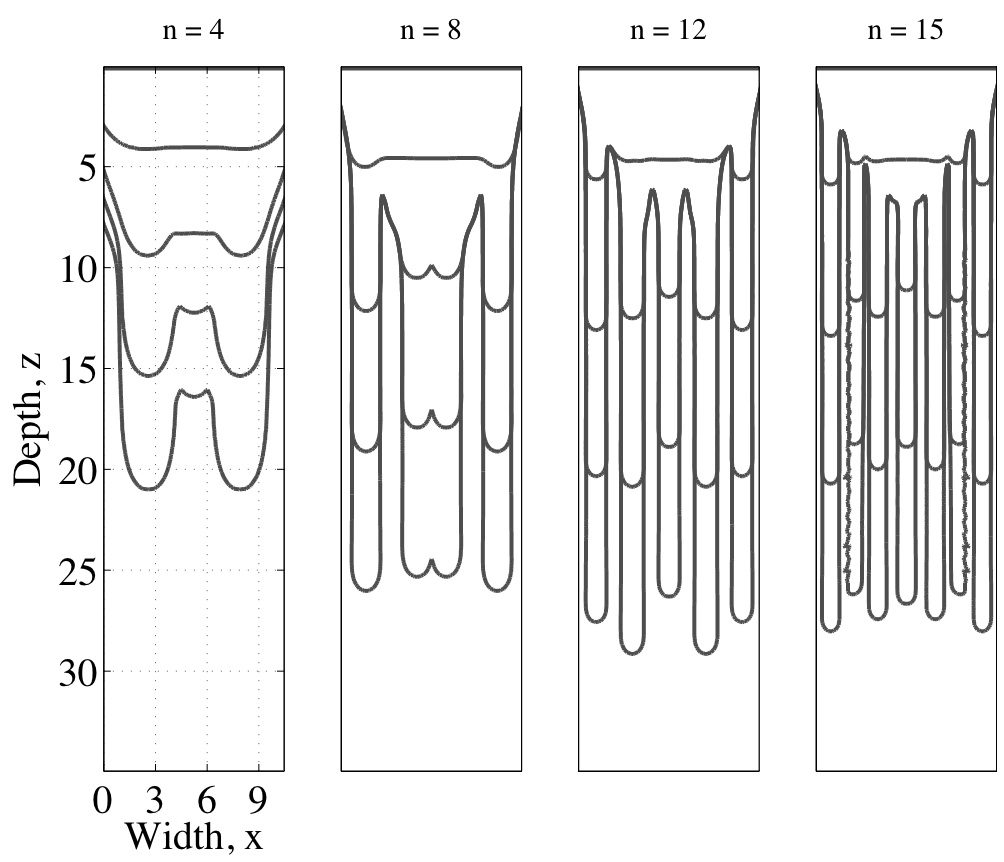}{n_sumplots.jpg}
  \caption{Contour plots of saturation for different values of the
    capillary shape parameter $n$, with contours shown at four
    equally-spaced times.  The base case is indicated by a 
    dagger ($\dag$).}  
  \label{fig:n}
\end{figure}

In Fig.~\ref{fig:n}, we present simulations for several values of $n$
lying between 4 and 15, while all other parameters are set to the base
case values.  In all simulations except for $n=4$, the solution exhibits
well-defined fingers having the characteristic non-monotonic saturation
profile down the central axis of each finger.  Furthermore, increasing
$n$ leads to an increase in the tip/tail saturation ratio and a decrease
in finger width and finger flux, as expected; in other words, fingers
are more diffuse for smaller $n$ while finger boundaries become sharper
when $n$ is increased.\ \  
The value of $n=12$ used in the base case is characteristic of highly
water-repellent soils, which is corresponds to the other experimental
and numerical studies we are most interested in. 

\subsection{Sensitivity to Initial Saturation, $\sat_i$}
\label{sec:sens-thetai}

A study of the effect of the initial saturation $\sat_i$ on our
numerical solution is warranted for two reasons.  First of all, the
nature of fingering instabilities and the properties of individual
fingers (such as finger width and velocity) can be very sensitive to the
choice of initial water content, as evidenced by several experimental
studies
\citep{diment-watson-1985,Bauters-etal-2000b,Wang-Tuli-Jury2003}.
Finger shape and size depends strongly on the initial wetting state; in
particular, vertical infiltration into a soil with larger initial
saturation tends to generate fingers that are more diffuse than when the
soil is dry.  Secondly, as with the capillary shape parameter $n$, the
value of $\sat_i$ is frequently omitted in the list of parameters
reported in experimental studies (e.g.,
\citet{Glass-Steenhuis-Parlange1989}).

\begin{figure}[hbtp]
  \centering  
  \psfrag{q}{}
  \psfrag{i}{}
  \psfrag{ = 0.001}{}
  \psfrag{ = 0.01}{}
  \psfrag{ = 0.03}{}
  \psfrag{ = 0.075}{}
  \replacewd
  {\mylabelsize
    \begin{tabular}{ccccc}
      & (a) & (b) & (c) & (d) \\[-0.3cm]
      & $\sat_i=0.001$ & \;$\sat_i=0.01^\dag$\; & 
      \;\;$\sat_i=0.03$\;\;  & \;$\sat_i=0.075$\; 
    \end{tabular}\\
  }
  \myfig{0.45\textwidth}{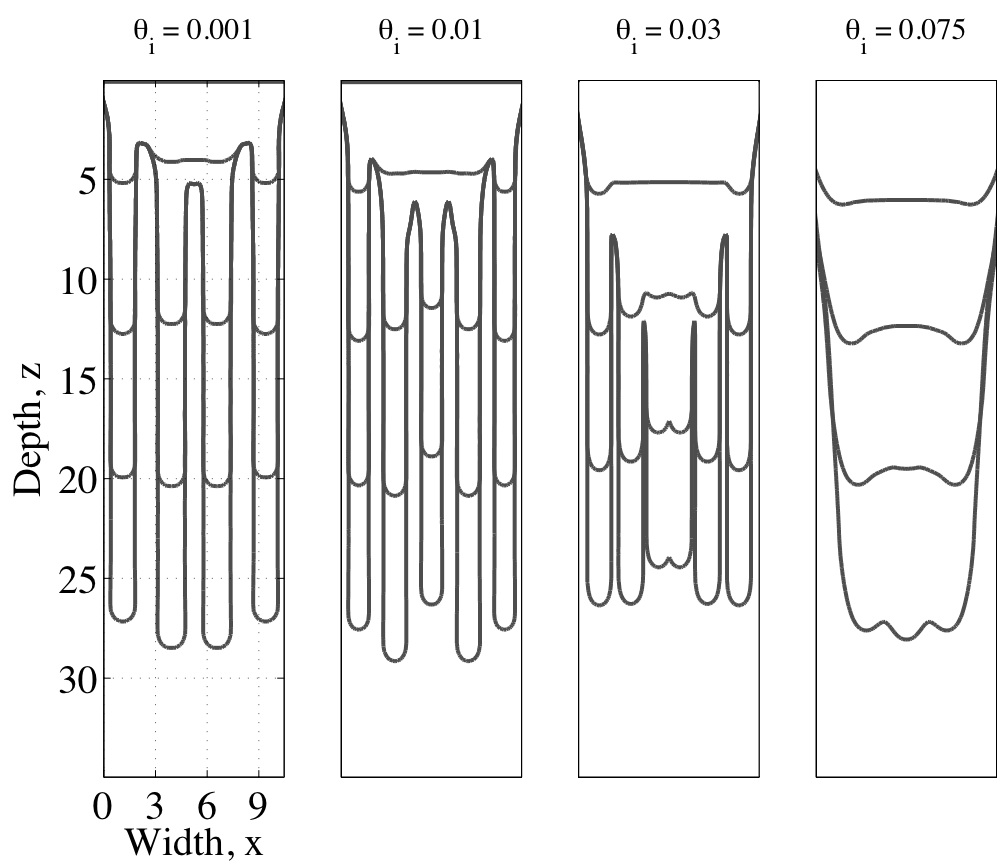}{sat_sumplots.jpg}
  \caption{Contour plots of saturation for different values of the
    initial saturation $\sat_i$, shown at four equally-spaced times.
    The base case is indicated by a dagger ($\dag$).}   
  \label{fig:sat}
\end{figure}

We therefore perform a series of simulations using various choices of
initial saturation between 0.001 and 0.075, holding the infiltration
flux $\flux_s$ constant.  The results are summarized in
Fig.~\ref{fig:sat} from which we observe that the number of fingers
increases as $\sat_i$ is increased.  In fact, the spacing between
fingers also decreases to the extent that when $\sat_i\gtrapprox 0.05$,
the individual fingers merge together to form a single finger.
\begin{figure}
  \centering  
  \psfrag{q}{}
  \psfrag{i}{}
  \psfrag{ = 0.001}{}
  \psfrag{ = 0.01}{}
  \psfrag{ = 0.03}{}
  \psfrag{ = 0.075}{}
  \replacewd
  {\mysmalllabelsize
    \begin{tabular}{ccccc}
      (a) & (b) & (c) & (d) & \\
      $\sat_i=0.001$ & $\sat_i=0.01^\dag$ & 
      $\sat_i=0.03$  & $\sat_i=0.075$ & \mbox{\qquad\quad}
    \end{tabular}\\[-0.0cm]
  }
  \myfig{0.5\textwidth}{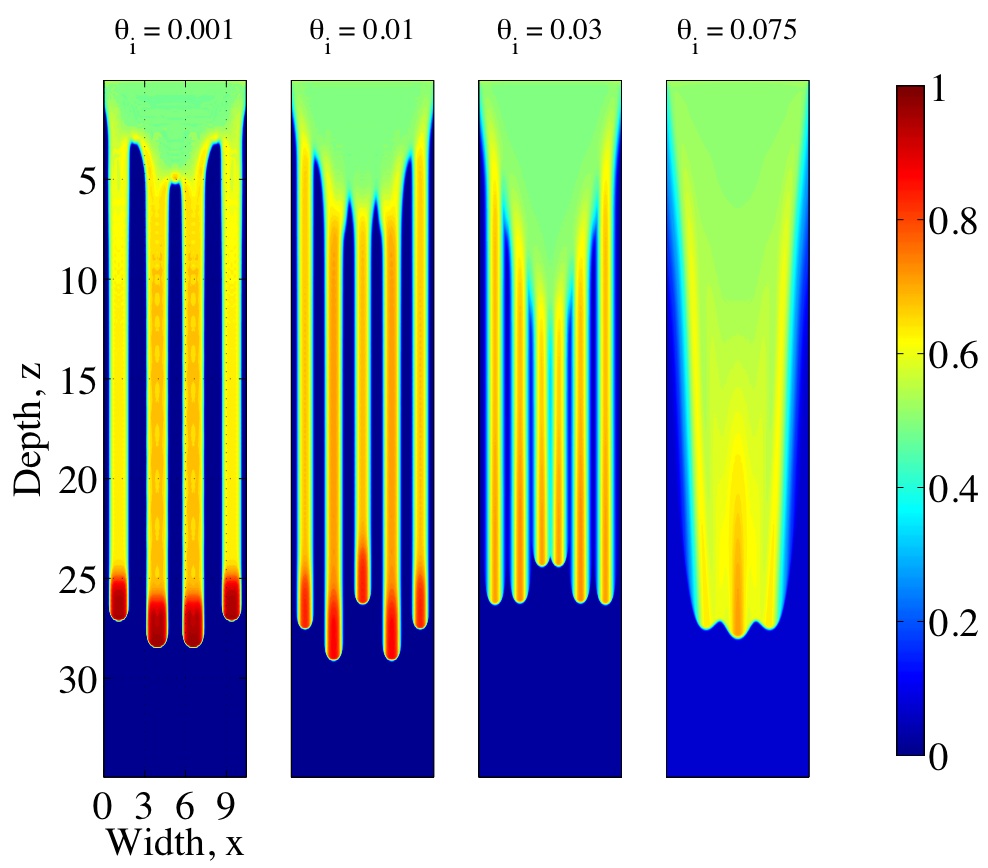}{sat_sumplots_bw.jpg}
  \caption{Saturation maps at the final time corresponding to the
    same values of $\sat_i$ depicted in {\protect Fig.~\ref{fig:sat}}.} 
  \label{fig:sat-map}
\end{figure}
As $\sat_i$ and finger width increase, we notice from the saturation
maps in Fig.~\ref{fig:sat-map} that the maximum finger tip saturation
decreases while the finger velocity remains relatively unchanged; this
behaviour can be justified using a simple mass conservation argument.
Most of these computed trends are consistent with experiments, the
exception being the finger velocity for which some experimental studies
exhibit a stronger dependence on $\sat_i$ (e.g.,
\citet{Bauters-etal-2000b}).

We mention in closing this section that in the absence of a given value of initial
saturation in \citet{Glass-Steenhuis-Parlange1989}, we have chosen
$\sat_i=0.01$ for the base case.  This value lies within with the
typical range of residual saturations seen in experiments for similar
soils, and also generates fingers with a tip saturation that is
consistent with values reported by \citet{DiCarlo2004}.

\subsection{Sensitivity to Capillary Relaxation Coefficient, $\tau_o$}
\label{sec:sens-tau}

Although several recent models for gravity-driven fingering have used a
capillary relaxation term to incorporate dynamic effects, there remains
a great deal of uncertainty in both the functional form and overall
magnitude of the relaxation coefficient
$\tau$~\citep{Juanes-2008,Manthey-etal2008}.\ \
\citet{Stauffer1978} derived an empirical estimate based on the
Brooks--Corey model for conductivity and capillary pressure (in lieu of
the van Genuchten--Mualem relationships used here) that takes the
following form
\begin{gather}
  \taus = \frac{\gamma_s\mu\sat_s h_b^2}{\lambda k_o}, 
  \label{eq:stauffer} 
\end{gather}
where $\gamma_s=0.1$ is a fitting parameter, $\mu$ is the fluid
viscosity, and $\lambda$ and $h_b$ are the Brooks--Corey parameters.
Recent experimental results suggest that the values of $\tau$ for sandy
media can range between 0.006 and 20, so that $\taus$ lies between
$2\times 10^3$ and $6\times 10^6\;kg/m\, s$ \citep{Manthey-etal2008}.

\begin{figure}[hbtp]
  \centering
  \psfrag{t}{}
  \psfrag{o}{}
  \psfrag{ = 0.01}{}
  \psfrag{ = 0.05}{}
  \psfrag{ = 0.1}{}
  \psfrag{ = 0.3}{}
  \replacewd
  {\mylabelsize
    \begin{tabular}{cccc}
      (a) $\tau_o=0.01$ & (b) $\tau_o=0.05$ & 
      (c) $\tau_o=0.1^\dag$ & (d) $\tau_o=0.3$
    \end{tabular}\\
  }
  \myfig{0.45\textwidth}{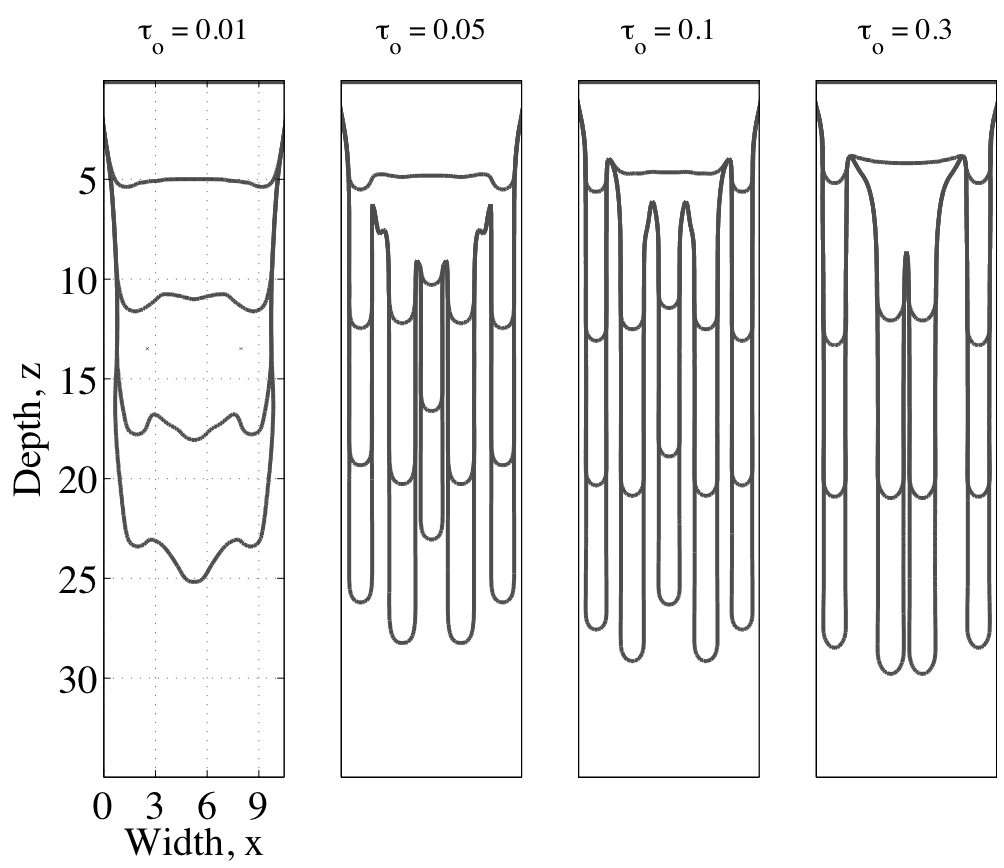}{tau_sumplots.jpg}
  \caption{Contours of saturation for different values of the capillary
    relaxation coefficient $\tau_o$, at four equally-spaced times.  The
    base case is indicated by a dagger ($\dag$).}
  \label{fig:tau}
\end{figure}

We have run a number of simulations using the functional form for $\tau$
given in \en{newrelax} and with the scaling constant $\tau_o$ varying
between 0.01 and 1.0.  The resulting saturation is depicted in
Fig.~\ref{fig:tau},\     
from which it is evident that $\tau_o$ has a strong influence not only
on the number of fingers but also on finger width and velocity.  If
$\tau_o$ is taken very small (less than 0.001) then dynamic effects
become negligible and finger formation is suppressed.  If, on the other
hand, $\tau_o$ is taken larger then the finger tip saturation tends to
increase which in turn reduces the number of fingers.

For the purposes of the base case, we have chosen an intermediate value
of $\tau_o=0.1$ which gives a range of $\tau$ that is centered on the
empirical estimate in Eq.~\en{stauffer}, and which also corresponds well
to the range of experimental values reported in the literature.

\subsection{Comparison with DiCarlo's Experiments} 
\label{sec:compare-dicarlo}

In this section we consider the experimental results reported by
\citet{DiCarlo2004}, who studied the importance of non-equilibrium
effects on finger formation in sandy porous media.  These experiments
investigated the effect of changes in infiltration flux, initial
saturation, and porous media properties on the resulting saturation
profiles.  \citeauthor{DiCarlo2005} also proposed an RE-based model
which neglected hysteretic effects, but did include a dynamic capillary
term (as in our Eqs.~\en{re}--\en{relax}) with a number of different
forms for the dynamic relaxation coefficient $\tau(\sat)$, including a
constant and various power-law forms similar to Eq.~\en{tau-psi}.
The correspondence between his numerical simulations and experiments (in
terms of finger tip saturation) was less than satisfactory; in
particular, although a reasonable fit was obtained for the tip
saturations when $\tau$ was a power law function, the wetting front
ahead of the finger tip was much too diffuse.
\begin{figure}[hbtp]
  \centering
  \psfrag{q}{}
  \psfrag{s}{}
  \psfrag{Saturation, }[cb][cB]{{\mylabelsize Saturation, $\sat$}}
  \psfrag{Infiltration flux, q}[cB][cb]{{\mylabelsize Infiltration flux, $\flux_s$}}
  (a) 20/30 sand\\
  \myfig{0.4\textwidth}{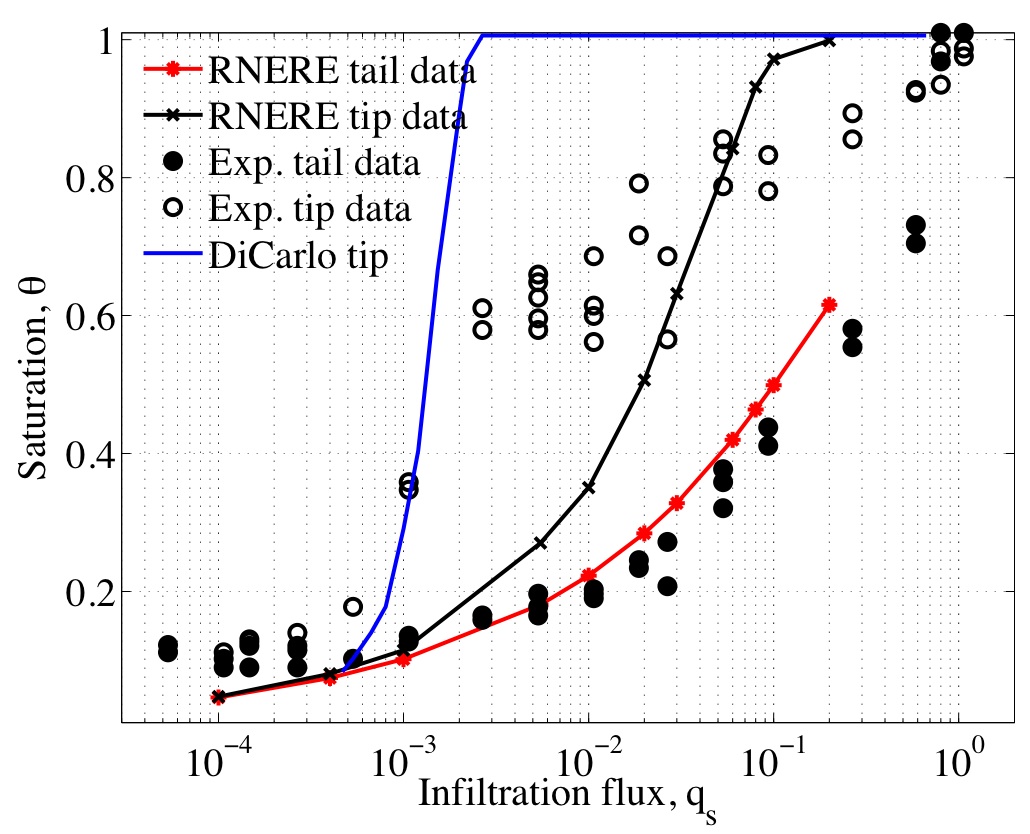}{dicarlo_tip_tail2030_mod.jpg}\\
  (b) 30/40 sand\\
  \myfig{0.4\textwidth}{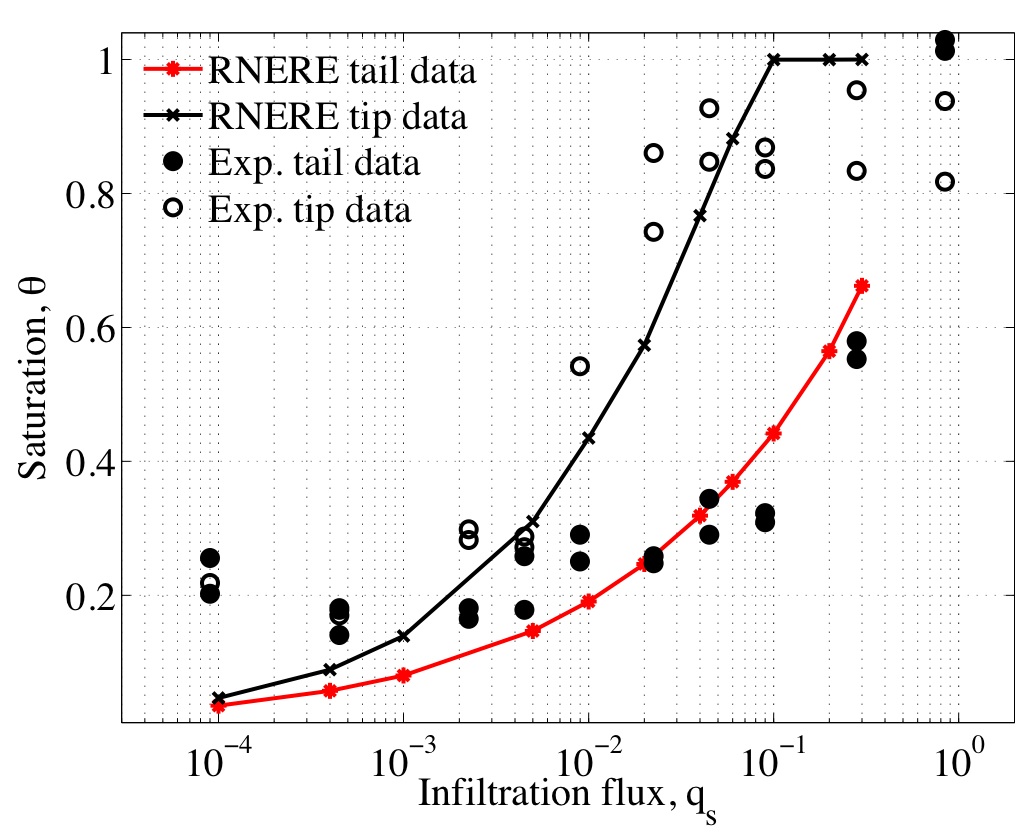}{dicarlo_tip_tail3040.jpg}
  \caption{Comparison of tip and tail saturations from the RNERE model
    and experiments.  Results are for two different soil types -- (a)
    20/30 sand (top), (b) 30/40 sand (bottom) -- and the experimental
    data points are extracted from \citet[Figs.~6 and~8]{DiCarlo2004}.
    The ``DiCarlo tip'' curve in (a) corresponds to a numerical
    simulation using a non-equilibrium model without hysteresis from
    \citet[Fig.~2]{DiCarlo2005}.}
  \label{fig:2030-3040} 
\end{figure}

We focus primarily on \citeauthor{DiCarlo2005}'s experimental and
numerical results for 20/30 sands which are reproduced in
Fig.~\ref{fig:2030-3040}(a).  These results were essentially
one-dimensional because the diameter of the soil columns being studied
was less than the characteristic finger width and hence was too small
for fingers to form; we have therefore performed a ``quasi-1D''
simulation in which the horizontal extent of the domain is only two grid
points wide.  Here, we choose parameters the same as in the base case
except that the capillary shape parameter, initial saturation, and
domain size are modified according to Table~\ref{tab:params-nond}(c).
Simulations were performed for a range of values of infiltration flux
$\flux_s$, and the resulting tip and tail saturations are plotted in
Fig.~\ref{fig:2030-3040}(a).  

We extracted values of all parameters from \citet{DiCarlo2004} except
for the initial saturation which he did not provide.  The computed tip
saturation is quite sensitive to $\sat_i$, while the impact on tail
saturation is much less.  As $\sat_i$ is increased, the tip profile in
Fig.~\ref{fig:2030-3040}(a) shifts toward the right until it eventually
overlaps with the tail profile, which corresponds to a stable flow.  On
the other hand, as $\sat_i$ is decreased the profile shifts to the left
and steepens becoming similar in shape to the ``DiCarlo tip'' curve.
Consequently, we have used initial saturation as a fitting parameter and
chose a value of $\sat_i=0.001$ that yields the best match with
experimental tip data.

The results in Fig.~\ref{fig:2030-3040}(a) demonstrate a significant
improvement over \citeauthor{DiCarlo2005}'s model, especially in terms
of the tip saturation.  There is also excellent agreement with the tail
data, although unfortunately there is no corresponding tail simulation
from \citeauthor{DiCarlo2005} for us to compare to.  A second comparison
is made for a 30/40 sand from \citet{DiCarlo2004} in
Fig.~\ref{fig:2030-3040}(b).  Except for some small deviations at the
lowest infiltration rates, these results also show a good fit between
our model and \citeauthor{DiCarlo2005}'s experiments.  Similar
comparisons are obtained for other soil types.

Finally, it is worth emphasizing that our computations exhibit fingers
that sustain a sharp front ahead of the finger tip, exhibiting none of
the non-physical diffusive smoothing observed in the model results of
\citet{DiCarlo2005}.  This discrepancy can be justified following
\citet{Nieber-Sheshukov2003} who attributed the initial formation of
fingers to dynamic capillary effects that are present in both DiCarlo's
and our RNERE model; however, fingers persist in time only when
hysteretic effects are also incorporated, which is the case for our
RNERE model but not \citeauthor{DiCarlo2005}'s.

\subsection{Comparison with Glass et al.'s Experiments}
\label{sec:compare-glass}  

We next make use of the dimensional analysis and experiments
of \citet{Glass-Parlange-Steenhuis1989,Glass-Steenhuis-Parlange1989} to
assess the response of finger width and tip velocity to changes in 
infiltration flux, $\flux_s$.\ \ 
Their two-dimensional experiments involved two layers of
fine-over-coarse sand, with water fed in from the top and air allowed to
escape freely.  Fingering was observed in the lower, coarse sand layer --
a result that can be predicted using the stability analysis of
\citet{Raats1973}.  We therefore restrict our attention to the lower
layer only which contains a coarse 14/20 silica sand having grain
diameter in the range 0.00070--0.0012 \units{m}.

The parameter values used in this section are the same as those for the
base case listed in Table~\ref{tab:params-nond}(a).  The residual
saturation $\sat_r=0.078$ is consistent with the tail water content
reported in \cite{Glass-Steenhuis-Parlange1989}; however, they did not
provide values of the remaining porous medium parameters and so we chose
$n=12$, $\sat_i=0.01$ and $\alphas_w=35\; m^{-1}$, which are consistent
with other 14/20 sands in the literature.

\begin{figure}[hbtp]
  \centering
  \psfrag{Q}{}
  \psfrag{s}{}
  \psfrag{Saturation, }[cb][cB]{{\mylabelsize Saturation, $\sat$}}
  \psfrag{Infiltration flux, q}[cB][cb]{{\mylabelsize Infiltration flux, $\flux_s$}}
  \myfig{0.4\textwidth}{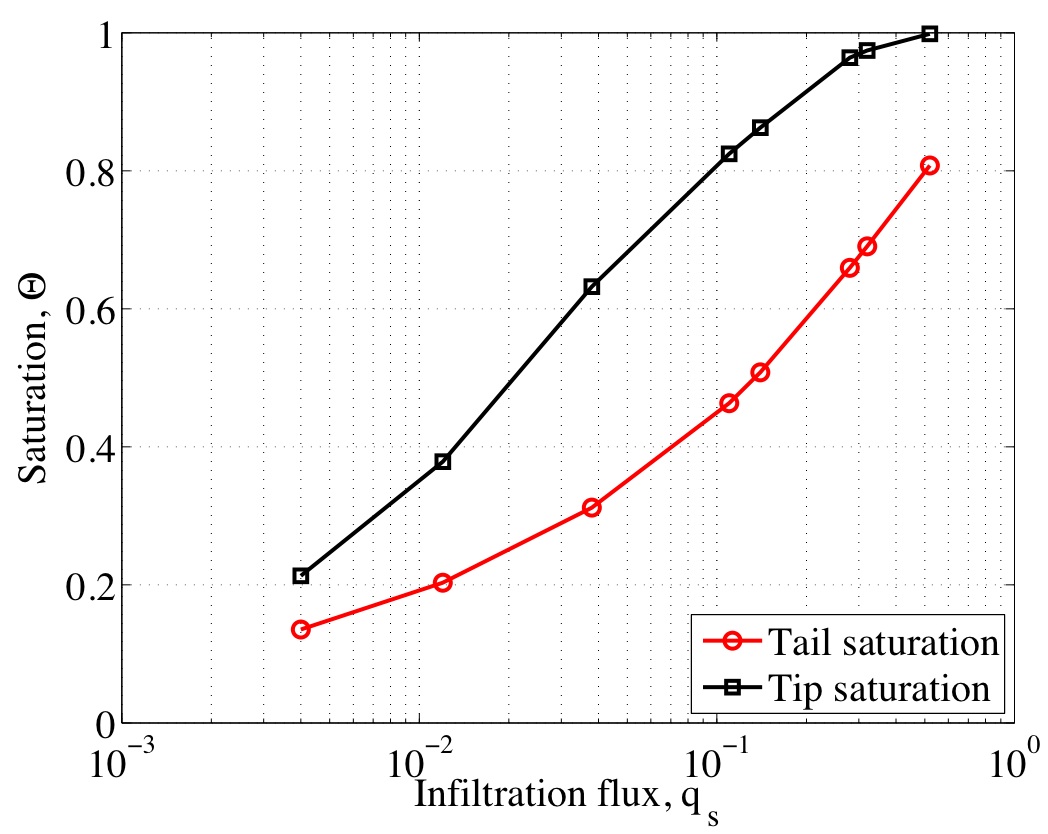}{tip_tail_plot.jpg}
  \caption{Computed tip and tail saturations for the
    \citet{Glass-Steenhuis-Parlange1989} comparisons.}
  \label{fig:tipglass}
\end{figure}

In Fig.~\ref{fig:glass_comp}, we present saturation contours from a
series of simulations in which the infiltration flux $\flux_s$ is varied
between 0.038 and 0.32~$cm/min$.  Decreasing the $\flux_s$ causes an
increase in the number of fingers, in addition to decreasing both finger
velocity and tip saturation (as indicated in Fig.~\ref{fig:tipglass}).
The corresponding numerical values for various quantities are summarized
in Table~\ref{tab:no_finger}.

\ifthenelse{\boolean{@IsSubmitted}}{\begin{figure}}%
  {\begin{figure*}}
  \def\ph{\phantom{age }}
  \def\pph{\phantom{th}}
  \newbox\dothis
  \setbox\dothis=\vbox to0pt{\vskip-1pt\hsize=11.5pc\centering
    Error\vss}
  \centering 
  \psfrag{s}{}
  \psfrag{q}{}
  \psfrag{ = 0.52}[ct][cb]{{\mylabelsize  (a) $\flux_s=0.52$}\quad}
  \psfrag{ = 0.32}[ct][cb]{{\mylabelsize  (b) $\flux_s=0.32$}\quad}
  \psfrag{ = 0.28}[ct][cb]{{\mylabelsize  (c) $\flux_s=0.28$}\quad}
  \psfrag{ = 0.14}[ct][cb]{{\mylabelsize  (d) $\flux_s=0.14^\dag$}\quad}
  \psfrag{ = 0.11}[ct][cb]{{\mylabelsize  (e) $\flux_s=0.11$}\quad}
  \psfrag{ = 0.088}[ct][cb]{{\mylabelsize (f) $\flux_s=0.088$}\quad}
  \psfrag{ = 0.038}[ct][cb]{{\mylabelsize (g) $\flux_s=0.038$}\quad}
  \psfrag{t = 42}[ct][cb]{{\mylabelsize $t=42$}}
  \psfrag{t = 54}[ct][cb]{{\mylabelsize $t=54$}}
  \psfrag{t = 58}[ct][cb]{{\mylabelsize $t=58$}}
  \psfrag{t = 77}[ct][cb]{{\mylabelsize $t=77$}}
  \psfrag{t = 84}[ct][cb]{{\mylabelsize $t=84$}}
  \psfrag{t = 96}[ct][cb]{{\mylabelsize $t=96$}}
  \psfrag{t = 153}[ct][cb]{{\mylabelsize $t=153$}}
  \replacewd
  \myfig{0.95\textwidth}{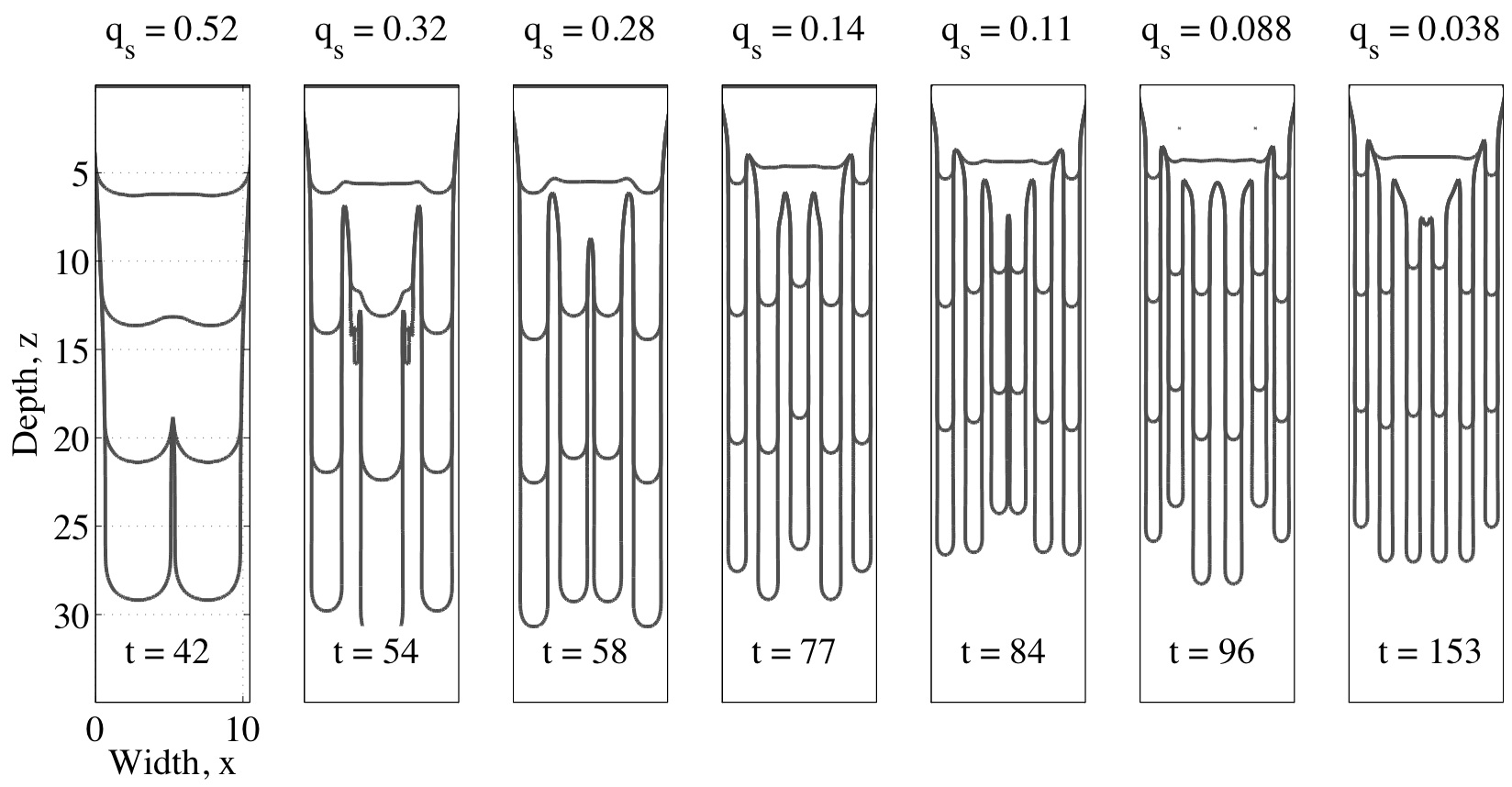}{glassall.jpg}
  \caption{Saturation contours for various values of infiltration flux
    $flux_s$, corresponding to parameters listed in
    Table~\ref{tab:params-nond}(a) for
    \citet{Glass-Steenhuis-Parlange1989}.  A further comparison of
    specific quantities is provided in Table~\ref{tab:no_finger}.  The
    base case is indicated by a dagger ($\dag$) and the end time for
    each simulation is indicated on the plot.}
    \label{fig:glass_comp}
\ifthenelse{\boolean{@IsSubmitted}}{\end{figure}}%
  {\end{figure*}}

\begin{table*}
  \def\ph{\phantom{age }}
  \def\pph{\phantom{th}}
  \newbox\dothis
  \setbox\dothis=\vbox to0pt{\vskip-1pt\hsize=11.5pc\centering
    Error\vss}
  \caption{Comparison of the finger number, width and velocity
    corresponding to the simulations in Fig.~\ref{fig:glass_comp}.  The
    experimental data are taken from
    \citet{Glass-Steenhuis-Parlange1989} (no data were available for
    the highest flux value, $\flux_s=0.52$).}    
  \label{tab:no_finger} 
  \ifthenelse{\boolean{@IsAnalyticalDf}}{
    \begin{tabular*}{0.99\textwidth}{@{\extracolsep{\fill}}c|cccc|cccc|cc}\hline 
      \\[-0.2cm]
      & \multicolumn{4}{c}{Experimental data}
      & \multicolumn{4}{c}{Numerical simulations}      
      & \multicolumn{2}{c}{Analytical estimates}
      \\ 
      \cline{2-5} \cline{6-9} \cline{10-11} \\
      $\flux_s$ & $N_f$ & $Q_f$ & $d_f$ & $v_f$
      & $N_f$ & $Q_f$ & $d_f$ & \hspace*{0.5cm}$v_f$\hspace*{0.5cm}
      & $d_f$--Eq.~\en{width1} & $d_f$--Eq.~\en{width2} \\[0.1cm]
      \hline
      0.012 & 4 & 0.03 & 0.45 & 0.12 & 7 & 0.02 & 0.28 & 0.08 & 0.60 & 1.19 \\
      0.038 & 4 & 0.10 & 0.52 & 0.21 & 6 & 0.07 & 0.34 & 0.18 & 0.62 & 1.20 \\
      0.088 & 5 & 0.18 & 0.60 & 0.29 & 6 & 0.15 & 0.49 & 0.29 & 0.65 & 1.23 \\
      0.11  & 6 & 0.19 & 0.61 & 0.30 & 6 & 0.19 & 0.54 & 0.32 & 0.67 & 1.25 \\
      0.14$\dag$ & 4 & 0.38 & 0.79 & 0.41 & 5 & 0.29 & 0.70 & 0.38 & 0.69 & 1.27 \\
      0.28  & 6 & 0.50 & 0.91 & 0.46 & 4 & 0.74 & 1.34 & 0.53 & 0.83 & 1.39 \\
      0.32  & 5 & 0.66 & 1.08 & 0.52 & 3 & 1.12 & 1.79 & 0.59 & 0.88 & 1.43 \\
      \hline
    \end{tabular*}
  }{
    \begin{tabular*}{0.99\textwidth}{*{9}{@{\extracolsep{\fill}}c}}\hline 
      \\[-0.2cm]
      & \multicolumn{4}{c}{Experimental data}
      & \multicolumn{4}{c}{Numerical simulations}      
      \\ 
      \cline{2-5} \cline{6-9} \\
      $\flux_s$ & $N_f$ & $Q_f$ & $d_f$ & $v_f$
      & $N_f$ & $Q_f$ & $d_f$ & \hspace*{0.5cm}$v_f$\hspace*{0.5cm}
      \\[0.1cm]
      \hline
      \hspace*{0.02\textwidth}0.012\hspace*{0.02\textwidth}
            & 4 & 0.03 & 0.45 & 0.12 & 7 & 0.02 & 0.28 & 0.08 \\
      0.038 & 4 & 0.10 & 0.52 & 0.21 & 6 & 0.07 & 0.34 & 0.18 \\
      0.088 & 5 & 0.18 & 0.60 & 0.29 & 6 & 0.15 & 0.49 & 0.29 \\
      0.11  & 6 & 0.19 & 0.61 & 0.30 & 6 & 0.19 & 0.54 & 0.32 \\
      0.14$\dag$ & 4 & 0.38 & 0.79 & 0.41 & 5 & 0.29 & 0.70 & 0.38 \\
      0.28  & 6 & 0.50 & 0.91 & 0.46 & 4 & 0.74 & 1.34 & 0.53 \\
      0.32  & 5 & 0.66 & 1.08 & 0.52 & 3 & 1.12 & 1.79 & 0.59 \\
      0.52  & --& --   & --   & --   & 2 & 2.73 & 3.88 & 0.69 \\
      \hline
    \end{tabular*}
  }
\end{table*}

\begin{figure}[hbtp]
  \centering
  \psfrag{f}{}
  \psfrag{Finger velocity, v}[cb][cB]{{\mylabelsize Finger velocity, $v_f$}}
  \psfrag{Flux through finger, Q}[cB][cb]{{\mylabelsize Finger flow rate, $Q_f$}}
  \psfrag{Exp. data}{{\scriptsize Exp. data}}
  \psfrag{Comp. data}{{\scriptsize Computed}}
  \myfig{0.4\textwidth}{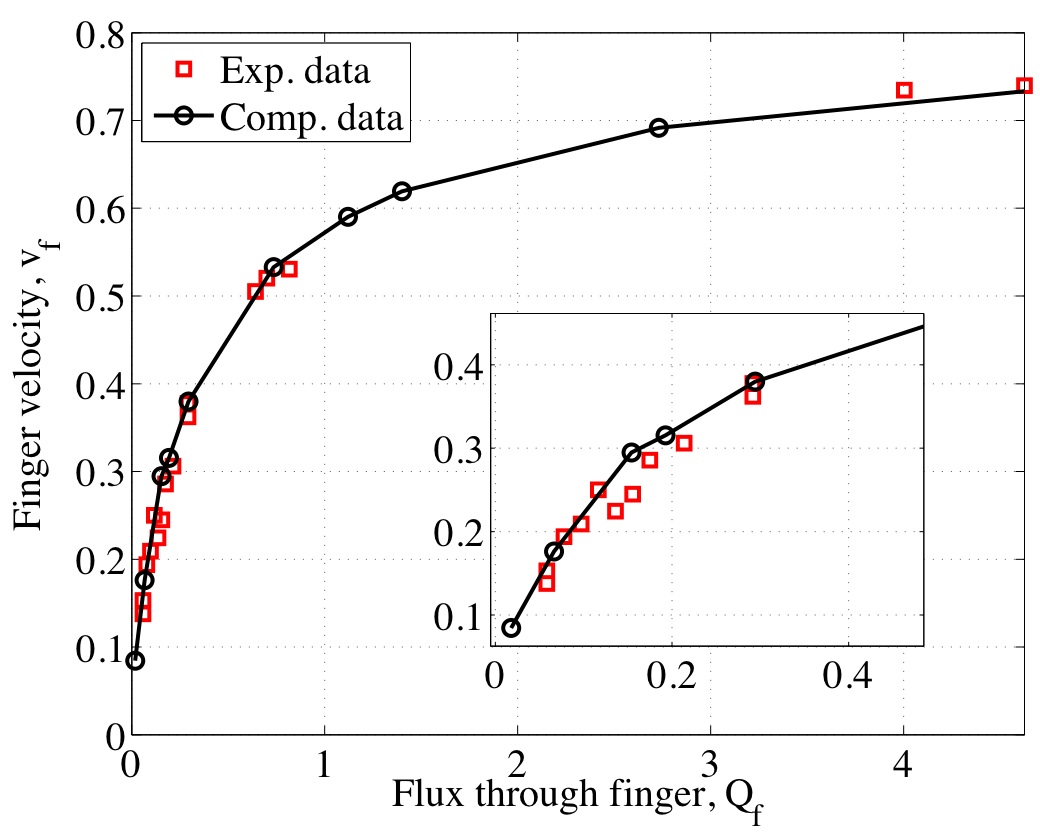}{flux_velocity_plot.jpg}
  \caption{Plot of finger flow rate versus velocity with experimental
    data (square points) taken from
    \citet{Glass-Steenhuis-Parlange1989}.}
  \label{fig:flowvel}
\end{figure}

Following \citet{Glass-Parlange-Steenhuis1989}, we relate the average
finger velocity ($v_f$) and width ($d_f$) to the average volume flow
rate in a finger using $Q_f = d_f \, v_f$.\ \
Since infiltration flux is related to finger velocity via $\flux_s = N_f
d_f v_f / d_i$, the finger volume flow rate can be written as
\begin{gather}
  Q_f = \frac{d_i \, \flux_s}{N_f}.
  \label{eq:flowrate}
\end{gather}
We then take our simulations for different values of $\flux_s$ and
redisplay the results in Fig.~\ref{fig:flowvel} as a plot of finger
velocity versus finger flow rate, including data from
\citeauthor{Glass-Steenhuis-Parlange1989}'s experiments.  There is very
close agreement between the simulated and experimental results.  In
particular, as the infiltration flux $\flux_s$ increases (or
equivalently, $v_f$ increases) the slope of the velocity--flux curve
decreases.

\begin{figure}[hbtp]
  \centering
  \psfrag{f}{}
  \psfrag{Finger width, d}[cb][cB]{{\mylabelsize Finger width, $d_f$}}
  \psfrag{Flux through finger, Q}[cB][cb]{{\mylabelsize Finger flow rate, $Q_f$}}
  \psfrag{Exp. data}{{\scriptsize Exp. data}}
  \psfrag{Comp. data}{{\scriptsize Computed}}
  \myfig{0.4\textwidth}{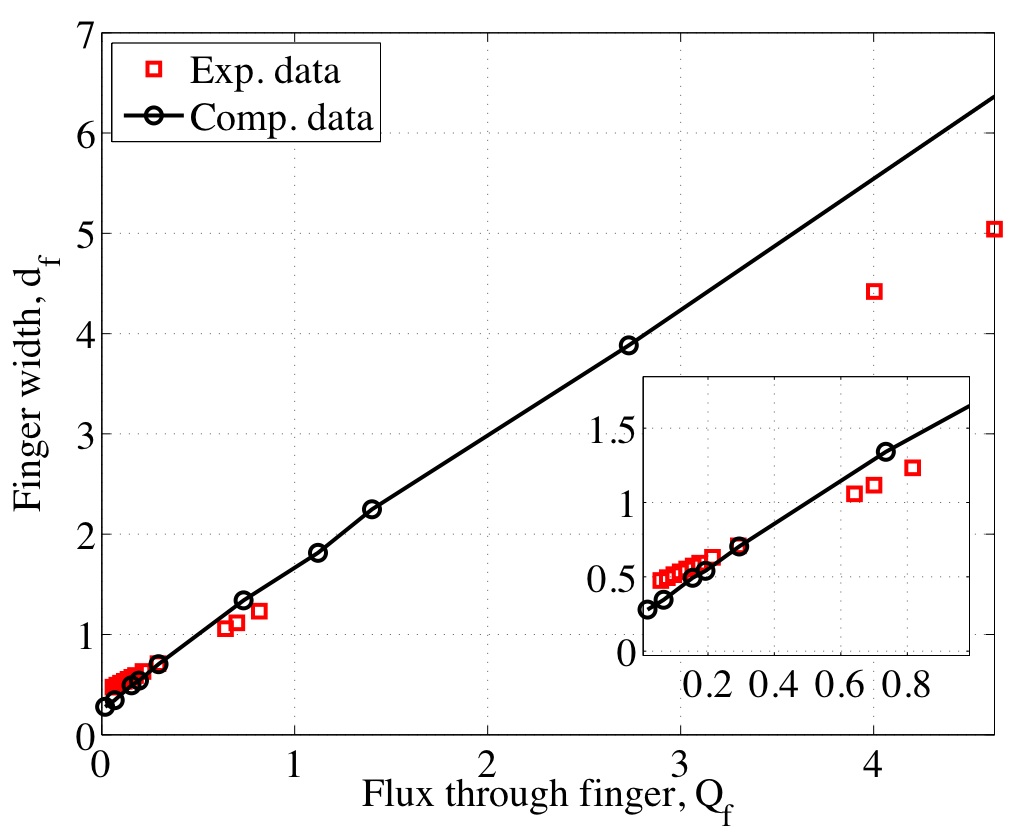}{flux_width_plot.jpg}
  \caption{Plot of flow rate versus finger width with experimental data
    (square points) taken from \citet{Glass-Steenhuis-Parlange1989}.}
  \label{fig:flowwidth}
\end{figure}

\begin{figure}[hbtp]
  \centering
  \psfrag{f}{}
  \psfrag{Finger velocity, q}[cb][cB]{{\mylabelsize Finger velocity, $v_f$}}
  \psfrag{Finger width, d}[cB][cb]{{\mylabelsize Finger width, $d_f$}}
  \myfig{0.4\textwidth}{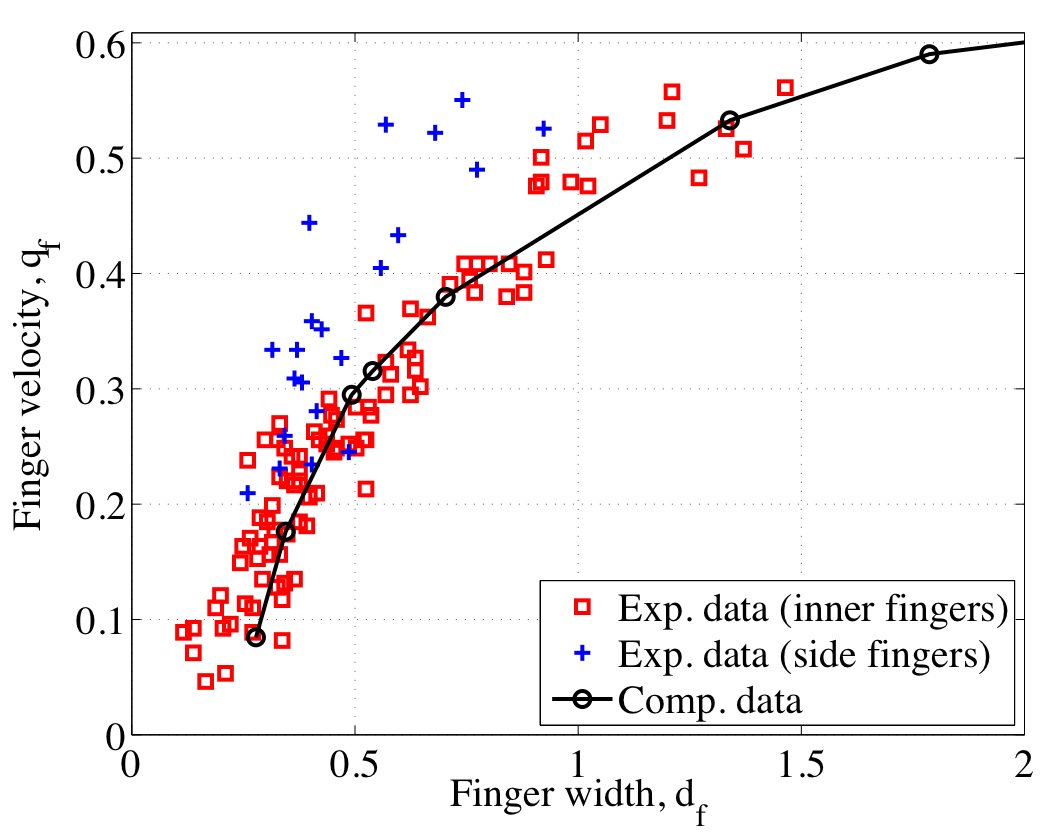}{vel_width_plot.jpg}
  \caption{Plot of finger width versus finger velocity, with
    experimental data taken from \citet{Glass-Steenhuis-Parlange1989}.
    Square data points denote fully developed interior fingers while
    crosses denote fingers adjacent to side boundaries.}
  \label{fig:widthvel}
\end{figure}

In Fig.~\ref{fig:flowwidth}, we present a plot of finger width versus
flow rate in which the dependence is approximately linear.  This
behavior is consistent with \citeauthor{Glass-Steenhuis-Parlange1989}'s
experiments where they used a linear least squares fit to predict the
finger width.  However, the correlation here is not as strong and our
computations significantly over-predict the finger width at higher
values of finger flux.
In an effort to explain this discrepancy, we plot finger velocity
against finger width in Fig.~\ref{fig:widthvel}, which includes the
experimental data of \citet{Glass-Steenhuis-Parlange1989}.  The
experimental points are classified as corresponding to ``side'' and
``inner'' fingers (where side fingers lie immediately adjacent to the
side boundaries) and significant differences are apparent between the
two sets of fingers which \citeauthor{Glass-Steenhuis-Parlange1989}
attribute to boundary effects.  If we focus only on the interior
fingers, then our model does a very good job of capturing the observed
behaviour.  Indeed, it is the contribution of the side fingers to the
average finger width that leads to the deviations in slope at higher
flux in Fig.~\ref{fig:flowwidth}.


\ifthenelse{\boolean{@IsAnalyticalDf}}{

  \subsection{Comparison with Analytical Approximations}
  \label{sec:compare-analytical}
  
  A number of analytical formulae have been derived in the literature that
  estimate the various finger properties (see Table~1 of
  \citet{Rooij2000}).  We consider one estimate
  \begin{gather}
    d_{f,1} = \frac{\pi S^2}{(\sat_s-\sat_i)}\left( \frac{1}{1-\flux_s}
    \right), \label{eq:width1} 
  \end{gather}
  based on sorptivity $S$ \citep{Parlange-Hill1976,Glass-etal1990}  and a
  second 
  \begin{gather}
    d_{f,2} = \pi\sqrt{a^*|\psiwe|}\left( \frac{1}{1-\flux_s}
    \right)^{1/2}, \label{eq:width2} 
  \end{gather}
  based on water entry pressure $\psiwe$ \citep{Wang-Feye1998}.  Both
  formulas are derived for the situation where the air at the top
  (infiltration) boundary is uncompressed and they are expressed in
  terms of our dimensionless variables.  Following
  \citet{Wang-Feye1998}, we set $a^*=rR^*\alpha_w$ where $r=1$ is the
  specific gravity and $R^*$ is a fitting parameter that depends on the
  wettability and microscale heterogeneity of the porous medium, which
  we simply take as $R^*=1$.  For
  \citeauthor{Glass-Steenhuis-Parlange1989}'s parameters, we present the
  analytical predictions for finger width in Table~\ref{tab:no_finger},
  which also lists the corresponding experimental and computed results
  plotted in Section~\ref{sec:compare-glass}.  Results in
  Table~\ref{tab:no_finger} affirm the good performance of the RNERE
  model especially at low values of infiltration flux.  However
  prediction of finger widths based on water entry pressure is very
  poor.  (THESE RESULTS ARE NOT GOOD AT ALL!  SINCE OTHER PAPERS ALSO
  SEEM TO HAVE SUCH A BAD MATCH TO THE ANALYTICAL FORMULAS, DO YOU SEE
  ANY REASON WHY WE SHOULD KEEP THIS?)

}{}

\section{Discussion}
\label{sec:discussion}

In this study we investigated the ability of the RNERE model to capture
gravity-driven fingering in unsaturated soils.  Our results expand on
previous studies of the RNERE model in two ways: first, by performing an
extensive sensitivity analysis for various important physical
parameters; and second, by drawing a systematic comparison between
simulations and previously published experimental results.  We showed
that with a careful choice of initial saturation and capillary
relaxation coefficient, the model is capable of accurately reproducing
the fingering behaviour observed in experiments.  Comparisons with
several independent experimental studies attest to the accuracy and
robustness of the RNERE approach.

In contrast to the work of \citet{DiCarlo2005}, who concluded that the
RNERE does not contain all the required physics to describe
gravity-driven fingering instabilities, we have shown that by coupling
both non-equilibrium and hysteretic effects it is possible to capture
fingering phenomena with the RNERE.  Our numerical simulations
demonstrate the importance of performing a detailed numerical
convergence study in order to ensure that fingers have been sufficiently
well resolved.  The model sensitivity analysis showed that dynamic
capillary terms must be properly handled if the fingered flow is to be
captured accurately, and in particular that an accurate estimate of the
$\tau_o$ parameter is essential.  As more research is undertaken in the
study of non-equilibrium capillary effects, we expect that more accurate
and reliable experimentally-validated correlations for the capillary
relaxation parameter will become available.

\ifthenelse{\boolean{@IsAnalyticalDf}}{
  The comparison between analytical expressions and experiments is very
  good and the slight discrepancies can be attributed to the following:
  \begin{itemize}
  \item most of these analytical expressions were derived for initially
    dry conditions and this will limit their applicability in comparison
    with experimental conditions. For example the expressions involving
    the sorptivity integral (see \citet{Parlange-Hill1976}) can only be
    evaluated accurately provided completely dry conditions are assumed.
  \item comparison of the RNERE model to \citet{DiCarlo2004} experiments
    showed that the finger tip saturation is sensitive to the initial
    saturation, see Fig \ref{fig:2030-3040}. 
  \item the numerical model and analytical expressions all assume
    isotropic conditions. This is difficult to control under experimental
    conditions (see for example \citet{Glass-Steenhuis-Parlange1989})
    where meandering of fingers might occur due to the heterogeneity of
    the porous media.
  \item In general, our numerical model performed far better because, for
    example \citet{Rooij2000} reported differences of 60 to 112\% between
    predicted and experimental finger widths. 
  \item we assumed closed loop hysteresis.
  \end{itemize}
}{}
  
There are a number of possible avenues for future work that will be
explored:
\begin{itemize}
\item We will investigate the use of alternate iterative strategies that
  improve on the robustness and efficiency of the RNERE algorithm.  We
  hope to draw inspiration in this respect from other well-known work on
  RE-based methods such as \citet{Celia1990} and
  \citet{Miller-etal-1998}.  Current advances in ODE solvers for dealing
  with event detection and non-smooth or discontinuous coefficients may
  also yield improvements in the treatment of hysteretic switching
  criteria, which has a big impact on convergence of the iterative
  scheme.  Furthermore, there has been an explosion of recent work on
  alternate models for handling dynamic capillary effects which could be
  applied here
  \citep{Beliaev-Schotting-2001,Sander-Glidewell-Norbury-2008,Peszynska-Yi-2008,Helmig-etal-2007,Manthey-etal2008}.

\item Analytical results for fingered flow, derived using asymptotic or
  other approximate methods, will be studied to gain a better
  understanding of the impact of hysteresis and dynamic effects on the
  mechanics of finger formation.  We will initially be guided by other
  previous work on traveling wave approximations for wetting fronts in
  the RNERE model with dynamic capillary effects
  \citep{DiCarlo2008,Nieber-Dautov2005} and hysteresis
  \citep{Dautov-Egorov2002,Sander-Glidewell-Norbury-2008}.


\item Two alternate mathematical models have recently been proposed that
  are relevant to capturing gravity-driven fingering phenomena.  The
  model of \citet{cuetofelgueroso-juanes-2008} accounts for effective
  surface tension phenomena due to saturation gradients through the
  addition of a new fourth order derivative term in the equations.
  Their numerical results capture the main qualitative features of
  fingered flow without the need for hysteresis, although they state
  that hysteretic effects are still important and that hysteresis can be
  easily incorporated into their model
  \citep{cuetofelgueroso-juanes-2009b}.  Another approach proposed by
  \citet{Pop-Etal2009} introduces an additional PDE for the interfacial
  area that obviates the need for an explicit treatment of hysteresis.
  We intend to perform an extensive computational comparison of these
  two models using a generalization of our RNERE approach, initially in
  1D, that should help to elucidate the relative advantages and
  disadvantages of the various approaches.
\end{itemize}

%
%

\begin{acknowledgments}
  This work was funded by research grants from the Natural Sciences and
  Engineering Research Council of Canada and the MITACS Network of
  Centres of Excellence.  JMS was supported by a Research Fellowship
  from the Alexander von Humboldt Foundation while visiting the
  Fraunhofer Institut Techno- und Wirtschaftsmathematik.  We are
  sincerely grateful to the anonymous referees whose extensive and
  thoughtful comments led to significant improvements in this paper.
\end{acknowledgments}

%
%

%
%

\ifthenelse{\boolean{@IsSubmitted}}{

}{

  \bibliographystyle{agu08}
  \bibliography{short-abbrevs,finger_bibil}

}

\end{article}

\end{document}